\documentclass[journal,draftclsnofoot,onecolumn,12pt]{IEEEtran}
\IEEEoverridecommandlockouts
\usepackage{amsmath,amsfonts}
\usepackage{algorithm}
\usepackage{array}
\usepackage[caption=false,font=normalsize,labelfont=sf,textfont=sf]{subfig}
\usepackage{textcomp}
\usepackage{stfloats}
\usepackage{url}
\usepackage{verbatim}
\usepackage{graphicx}
\usepackage{cite}
\usepackage{booktabs} 
\usepackage{threeparttable} 
\usepackage{pifont} 

\IEEEoverridecommandlockouts
\usepackage{algorithmicx}
\usepackage{pdfcomment}
\usepackage{bookmark}
\usepackage{enumerate}
\usepackage{bm}
\usepackage{amsmath,epsfig,amssymb,amsfonts,color}

\newcommand{\reffig}[1]{Fig. \ref{#1}}

\usepackage{multirow}

\newcommand{\tabincell}[2]{\begin{tabular}{@{}#1@{}}#2\end{tabular}}  

\begin{document}

\title{Quantized Phase Alignment by Discrete Phase Shifts for Reconfigurable Intelligent Surface-Assisted Communication Systems}

\author{Jian Sang, Jifeng Lan, Mingyong Zhou, Boning Gao, Wankai Tang, Xiao Li, Xinping Yi, and Shi Jin
\thanks{Jian Sang, Jifeng Lan, Mingyong Zhou, Boning Gao, Wankai Tang, Xiao Li, and Shi Jin are with the National Mobile Communications Research Laboratory, Southeast University, Nanjing 210012, China.}
\thanks{Xinping Yi is with the Department of Electrical Engineering and Electronics, University of Liverpool, Liverpool L69 3BX, U.K.}

}




\maketitle

\begin{abstract}
Reconfigurable intelligent surface (RIS) has aroused a surge of interest in recent years. In this paper, we investigate the joint phase alignment and phase quantization on discrete phase shift designs for RIS-assisted single-input single-output (SISO) system. Firstly, the phenomena of phase distribution in far field and near field are respectively unveiled, paving the way for discretization of phase shift for RIS. Then, aiming at aligning phases, the phase distribution law and its underlying degree-of-freedom (DoF) are characterized, serving as the guideline of phase quantization strategies. Subsequently, two phase quantization methods, dynamic threshold phase quantization (DTPQ) and equal interval phase quantization (EIPQ), are proposed to strengthen the beamforming effect of RIS. DTPQ is capable of calculating the optimal discrete phase shifts with linear complexity in the number of unit cells on RIS, whilst EIPQ is a simplified method with a constant complexity yielding sub-optimal solution. Simulation results demonstrate that both methods achieve substantial improvements on power gain, stability, and robustness over traditional quantization methods. The path loss (PL) scaling law under discrete phase shift of RIS is unveiled for the first time, with the phase shifts designed by DTPQ due to its optimality. Additionally, the field trials conducted at 2.6 GHz and 35 GHz validate the favourable performance of the proposed methods in practical communication environment. 
\end{abstract}


\section{Introduction}
\label{sec1}


\IEEEPARstart{R}{econfigurable} intelligent surface (RIS), due to its ability of customizing the propagation environment controllably, has been widely considered as one of the revolutionary techniques for the sixth generation (6G) wireless communication systems \cite{b2, v2_b2}. 
RIS provides many innovative communication paradigms such as beamforming by co-phasing the signals reflected by it, which enable to improve the harsh communication link and even redefine the radio environments \cite{ v2_b3}. 

Generally, RIS is referred to a two-dimensional meta-surface, where plentiful unit cells are embedded and arranged periodically \cite{b1, v2_b6}. Though appropriate voltage control, these unit cells can tune the characteristics of electromagnetic (EM) waves impinging onto RIS, such as phase, to reconfigure the spatial phase gradient. In this way, a single shaped beam or even multiple shaped beams can be formed towards the desired direction flexibly \cite{b3}.


Presently, RIS has been studied in many fields including but not limited to: index modulation \cite{b_s5}, environmental sensing and positioning \cite{b_s4}, and secure communications \cite{b_s2,b_s3}. Compared to traditionally used relaying, RIS requires negligible power consumption while achieving considerable transmission quality improvement, which agrees well with the claims of ``Green Communications'' \cite{v2_b7}. In addition, considering the plentiful unit cells on RIS as tiny antennas, RIS can be regarded as a low-cost and energy-efficient alternative to large-scale multiple-input multiple-output (MIMO) systems to perform beamforming \cite{v2_b8}. Undoubtedly, state-of-the-art RIS-assisted systems rely on phase shift designs, which is of great significance for improving the RIS-empowered channel quality, while its digitization faces some indispensable challenges in practice.  


\subsection{Related Works}

Up to now, numerous algorithm designs and characteristic analyses on continuous phase shifts of RIS have been proposed \cite{b4,b5,b_r8,v2_b1}. In \cite{b4} and \cite{b5}, the authors formulated the ideal continuous phase shifts of RIS based on the wave-path distances of EM waves. By feat of the deep reinforcement learning (DRL) framework, \cite{b_r8} optimized the continuous phase shift issues to maximize the downlink received signal-to-noise ratio (SNR) for RIS-assisted multiple-input single-output (MISO) system, with relatively low running time. Utilizing the fractional programming and complex circle manifold, \cite{v2_b1} investigated the continuous RIS phase shift designs under statistical channel state information for RIS-assisted multi-cell MISO communications.

Meanwhile, for practical deployment and application, the RIS with discrete phase shifts is still the focus in the related field \cite{b6,b13,b_r1,b_r3,b_r5,b7,b8,b9,b10,b11}. By appropriately configuring discrete phase shifts of RIS, \cite{b6} showed that the multi-path fading channel could be mitigated. In \cite{b13}, it was shown that a constant proportional power loss was incurred by discrete phase shifts, as compared to ideal continuous phase shifts. A hybrid beamforming scheme, which employed continuous beamforming at base station (BS) and discrete beamforming at RIS, was proposed in \cite{b_r1} for RIS-assisted downlink multiuser system to maximize the sum-rate. In \cite{b_r3}, the authors proposed a RIS-assisted MIMO framework where the discrete phase shifts of RIS were pre-designed and the effective superposed channel was estimated instead of separately training the BS-UE and BS-RIS-UE links. It was revealed in \cite{b_r5} that the minimum required number of discrete phase quantization levels is 3-bit, in order to achieve the full diversity order in RIS-assisted communication systems. 


The above-mentioned studies enriched the related research on phase shift designs of RIS, yet the discretization of phase shift as well as its transformation metrics from continuous phase shift, i.e., phase alignment and phase quantization, still receive rare attention. While the perfect phase alignment in continuous space can be easily achieved, it is challenging in discrete space due to the fact that phase alignment is highly dependent of the quantization strategies. In particular, as a many-to-few mapping, phase quantization is usually a non-linear and irreversible operation, which relies critically on the phase distribution of EM waves. Therefore discrete phase alignment should also take both phase distribution and quantization strategies into account.

In addition, the path loss (PL) scaling law for RIS-assisted channel, which is also inseparable from phase shift designs, acts as one crucial role in RIS-related research. In existing works, such as \cite{b4, b5, b_r9, b_r10, b_r11, b_r12, b_r13, b_r14}, the authors formulated that the PL scaling law in the far field is proportional to $(d_1d_2)^2$, where $d_1$ and $d_2$ are the distance from transmitter (Tx) and receiver (Rx) to the center of RIS, respectively. However, the existing studies were mainly analyzed under ideal continuous phase shifts of RIS. Under practical discrete phase shifts, the RIS-related PL characteristics have not been reported analytically, although it may be of great significance in many RIS-related fields such as channel modeling.

\subsection{Main Contributions}

Motivated by these circumstances, in this paper, we investigate the phase alignment and phase quantization for discrete phase shift design in RIS-assisted single-input single-output (SISO) channel. For the convenience of presentation, we refer to the cases of continuous and discrete phase shifts as ``continuous space'' and ``discrete space'', respectively. The contributions of this paper are summarized as follows. 


\begin{itemize}

\item By showcasing the phase distribution of EM waves, we characterize the properties of phase alignment in far field and near field respectively for RIS-assisted SISO systems. Aiming to maximize the beamforming effect, we reveal and prove the phase distribution law and its underlying degree-of-freedom (DoF), which serve as the guideline for phase quantization strategies. It is explained that the DoF of optimal solution for discrete phase shifts of RIS is proportional to the number of unit cells linearly, rather than the well-known exponentially. It is also pointed out that the DoF is only related to the number of unit cells and irrelevant to the number of bits. Accordingly, the optimal discrete phase shifts of RIS can be identified from quantizing the ideal continuous counterparts with quantization thresholds appropriately set.
 
\item In order to calculate the desired discrete phase shifts of RIS, we propose two phase quantization methods, i.e., dynamic threshold phase quantization (DTPQ) and equal interval phase quantization (EIPQ). Both of them are suitable for configuring discrete phase shifts with an arbitrary number of bit(s). DTPQ method is capable of obtaining the optimal discrete phase shifts. It has a complexity of $O(N_R)$, which is significantly reduced compared to the exhaustive search method, whose complexity is $O(2^{qN_R})$, where $N_R$, $q$ are the numbers of unit cells and bits respectively. Moreover, EIPQ is a sub-optimal method yet further reduces the computation complexity of DTPQ from linear to constant magnitude, while still showing a better beamforming performance than traditional quantization method.

\item The simulation results manifest that the proposed methods are capable of providing a more significant beamforming gain against the traditional quantization methods. We also diagrammatize the comparison on computational complexity of the proposed methods, the traditional quantization method, and the exhaustive search method. Moreover, the proposed methods demonstrate a more favourable robustness. Based on the phase shifts designed by DTPQ, the PL scaling law in discrete space for RIS-assisted SISO channel is analyzed for the first time. Noteworthily, it is found that the dependence of PL on angles of arrival/departure in discrete space is slightly different from that in continuous space. 

\item Field trials at sub-6 GHz and millimeter wave (mmWave) frequency bands are carried out respectively, to validate the beamforming performance of the proposed method in real communication environment. The trial results indicate a well agreement with the theoretical analyses and simulation results.

\end{itemize}

\subsection{Organization}

The remainder of this paper is structured as follows. Section II describes the model of RIS-assisted system and formulates the joint problem of phase alignment and phase quantization. In Section III, we showcase the phase distribution of EM waves and characterize the properties of phase alignment. Then the phase distribution law and its DoF are unveiled in Section IV. Two phase quantization methods are also proposed in this section. Subsequently in Section V, simulations are represented to compare the performances of the proposed methods against the traditional methods. Moreover, the PL scaling laws in discrete space and several field trials are also illustrated in this section. Finally, we conclude this paper in Section VI.

\section{System Model and Problem Formulation}
\label{sec2}
\subsection{System model}

\begin{figure}[htbp]
\centering
\includegraphics[width=6in]{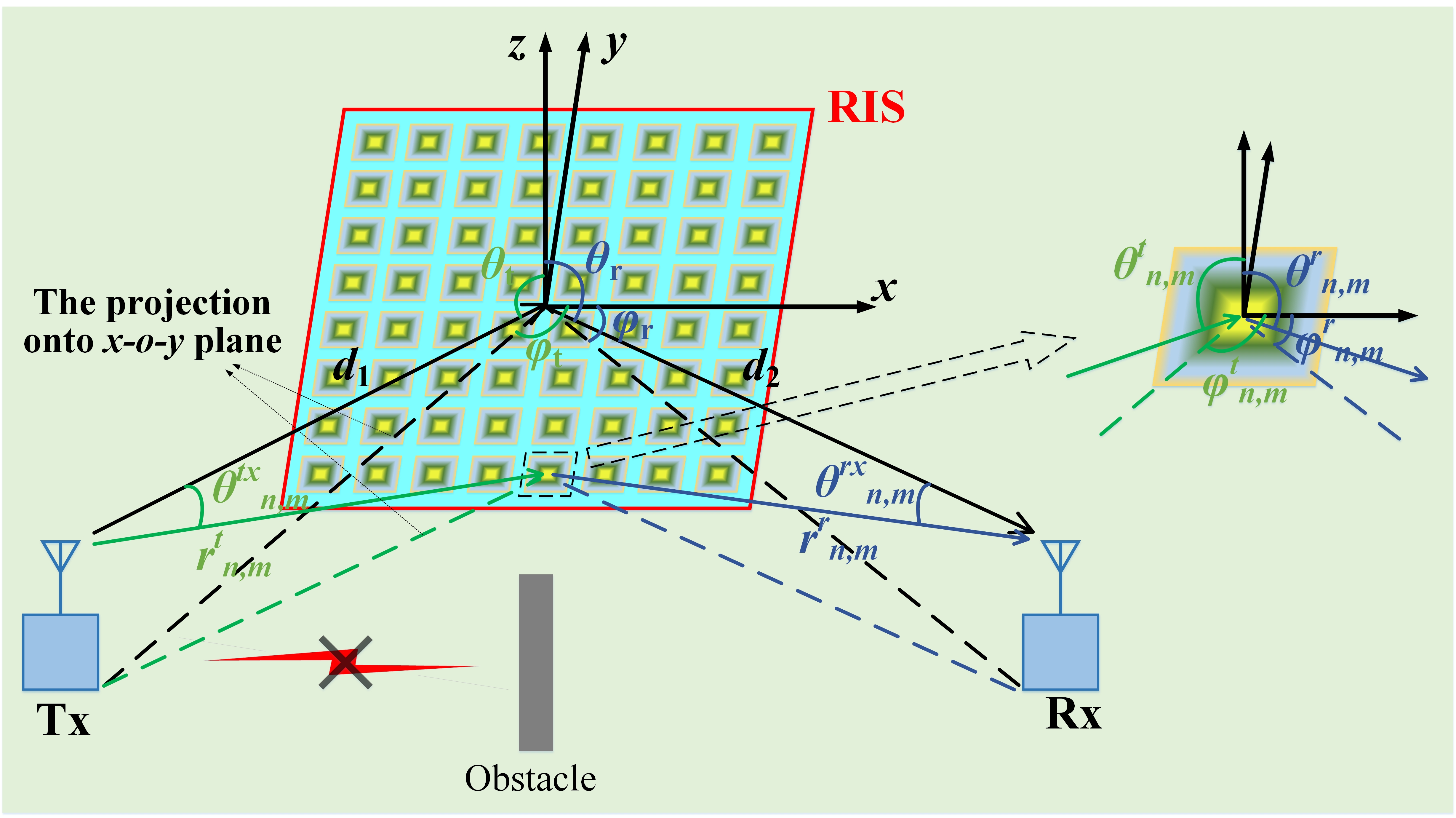}
\caption{RIS-assisted system model.}
\label{fig1}
\vspace{-0.2cm}
\end{figure}

As shown in \reffig{fig1}, a RIS-assisted SISO communication system is considered. Note that the RIS in this paper refers to the family of reflective RIS, by which the signal is neither amplified nor retransmitted. The direct channel between Tx and Rx is considered to be completely blocked and only the RIS-assisted cascaded channel is taken into account in the rest of this paper. The RIS consists of $M$ unit cells per row and $N$ unit cells per column, i.e., $M \times N$ unit cells in total. Assume that both of $M$ and $N$ are even numbers. Each unit cell has a rectangular shape with a size of $d_x$ along the $x$ axis and $d_y$ along the $y$ axis, as shown in \reffig{fig1}. Moreover, the reflection magnitude of unit cell is constant and its phase can be programmed with a $q$-bit resolution, i.e., ${2^q}$ phase shift levels in total. Let $d_1$ and $d_2$ refer to the distance from Tx to the center of RIS and the distance from Rx to the center of RIS, respectively. ${\theta_t}$, ${\varphi_t}$, ${\theta_r}$ and ${\varphi_r}$ indicate the elevation angle of arrival (EoA) and the azimuth angle of arrival (AoA) from Tx to the center of RIS as well as the elevation angle of departure (EoD) and the azimuth angle of departure (AoD) from the center of RIS to Rx. 

Denote the unit cell in the $n$th row and $m$th column as $\mathrm{U}_{n,m}$, then ${r_{n,m}^t}$, ${r_{n,m}^r}$, ${\theta _{n,m}^t}$, ${\varphi _{n,m}^t}$, ${\theta _{n,m}^r}$ and ${\varphi _{n,m}^r}$ represent the distance between Tx and $\mathrm{U}_{n,m}$, the distance between Rx and $\mathrm{U}_{n,m}$, the EoA and the AoA from Tx to $\mathrm{U}_{n,m}$, the EoD and the AoD from $\mathrm{U}_{n,m}$ to Rx, respectively. Considering $\theta _{n,m}^{tx}$ and $\varphi _{n,m}^{tx}$ denote the EoD and the AoD from Tx to $\mathrm{U}_{n,m}$ as well as $\theta _{n,m}^{rx}$ and $\varphi _{n,m}^{rx}$ represent the EoA and the AoA from $\mathrm{U}_{n,m}$ to Rx, respectively. Assume that there is always matched polarization among Tx, RIS, and Rx, then the received power at Rx can be expressed as follows \cite{b5}.
\begin{equation}
\label{eq1}
{P_r} = {P_t}\frac{{{G_t}{G_r}{{({d_x}{d_y})}^2}{A^2}}}{{16{\pi ^2}}} { \xi }^2,
\end{equation}
where $P_t$ denotes the transmitted power from Tx, $A$ denotes the reflection magnitude of $\mathrm{U}_{n,m}$, which is considered to be 1 in this paper, $\xi$ denotes the superposition effect of EM field at Rx and can be expressed as \cite{b5}
\begin{equation}
\label{eq2}
{ \xi(\bf{\Phi}) }= {\left| {\sum\limits_{m = 1}^M {\sum\limits_{n = 1}^N {\frac{{\sqrt {F_{n,m}^{combine}} }}{{r_{n,m}^tr_{n,m}^r}}{e^{ - j\left( {\frac{{2\pi (r_{n,m}^t + r_{n,m}^r)}}{\lambda } - {\phi _{n,m}}} \right)}}} } } \right|},
\end{equation}
where $|\cdot|$ represents the modulo operation, $\lambda$ is wavelength of operating frequency, $\bf{\Phi}$ denotes the phase shift matrix of RIS and ${\phi _{n,m}}$ is the phase shift of $\mathrm{U}_{n,m}$, which can be configured artificially. Additionally, $F_{n,m}^{combine}$ represents the joint effect of the normalized power radiation patterns and can be expressed as 
\begin{equation}
\label{eq_r1}
\begin{aligned}
F_{n,m}^{combine} = & {F^{tx}}(\theta _{n,m}^{tx},\varphi _{n,m}^{tx})F(\theta _{n,m}^t,\varphi _{n,m}^t)F(\theta _{n,m}^r,\varphi _{n,m}^r){F^{rx}}(\theta _{n,m}^{rx},\varphi _{n,m}^{rx}),
\end{aligned}
\end{equation}
where ${F^{tx}}(\theta _{n,m}^{tx},\varphi _{n,m}^{tx})$, $F(\theta _{n,m}^t,\varphi _{n,m}^t)$, $F(\theta _{n,m}^r,\varphi _{n,m}^r)$ and ${F^{rx}}(\theta _{n,m}^{rx},\varphi _{n,m}^{rx})$ denote the normalized power radiation patterns of Tx, emission of $\mathrm{U}_{n,m}$, reception of $\mathrm{U}_{n,m}$, and Rx respectively, which could be defined in the form of \eqref{eq_r2} \cite{b5}.
\begin{equation}
\label{eq_r2}
\begin{aligned}
F(\theta ,\varphi ) = \left\{ {\begin{array}{*{20}{l}}
{\left( \cos \theta  \right)^\alpha, } \quad   &\theta  \in [ 0, \frac{\pi}{2} ), \varphi \in [0,2\pi) ,\\
0, &\theta  \in [ {\frac{\pi }{2},{\rm{ }}\pi } ),\varphi \in [ 0,2\pi).
\end{array}} \right.
\end{aligned}
\end{equation}

In addition, $G_t$ and $G_r$ are the antenna gains of Tx and Rx respectively, which can be defined in the form of \eqref{eq_r3} \cite{b5}.
\begin{equation}
\label{eq_r3}
\begin{array}{l}
\begin{aligned}
G  = \frac{{4\pi }}{{\int\limits_{\varphi  = 0}^{2\pi } {\int\limits_{\theta  = 0}^\pi  {F(\theta ,\varphi )\sin \theta d\theta d\varphi } } }}
  = 2\left( {\alpha  + 1} \right).
\end{aligned}
\end{array}
\end{equation}
For the normalized power radiation patterns of Tx and Rx, $\alpha$ could be inferred from their gain $G_t$ and $G_r$ according to \eqref{eq_r3}. For the normalized power radiation patterns of unit cell on RIS, $\alpha$ is considered to be 1 in this paper.

\subsection{Phase quantization and alignment}

As we can see from \eqref{eq1} and \eqref{eq2}, when the positions of Tx, Rx and RIS are fixed, all of the parameters in \eqref{eq1} and \eqref{eq2} are constant values, except for the phase shift matrix $\bf{\Phi}$. Therefore, the configuration of phase shift matrix on RIS is of great significance for enhancing beamforming effect. A perfect phase alignment in continuous space is to align the phases in all $MN$ components of \eqref{eq2}, so that the magnitude of $\xi$ is maximized. Given the perfect knowledge of RIS-assisted channel in \reffig{fig1}, the wave-path distance of Tx-RIS-Rx link can be known. Then the ideal continuous phase shift matrix $\bf{\Phi}$ achieving perfect phase alignment can be readily obtained when its element $\phi_{n,m}$ in the $n$th row and $m$th column is designed as follows \cite{b4}.
\begin{equation}
\label{eq_r5}
\begin{array}{c}
{\phi_{n,m}} = \bmod \left({2\pi (r_{n,m}^t + r_{n,m}^r)}/{\lambda },~2\pi \right), \forall \ m \in \left\{ {1,2,...,{M}} \right\}, \forall \ n \in \left\{ {1,2,...,{N}} \right\}.
\end{array}
\end{equation}

However, due to hardware limitations, it is preferred to configure discrete phase shifts of RIS in practice \cite{b7,b8,b9,b10,b11}, which requires quantizing the above continuous phase shifts. This means that for the expressions in \eqref{eq2} in practical communication environment, the parameter $\bf{\Phi}$ should be correspondingly replaced by discrete phase shift matrix $\bf{\Delta}$, where its element in the $n$th row and $m$th column is denoted as $\Delta _{n,m}$.

Consider $q$-bit phase resolution with a uniform phase quantization interval $\Omega= 2\pi/2^q$, such that $\phi_{n,m}$ can be quantized by $2^q$ uniformly distributed discrete phase shift levels,
\begin{equation}
\label{eq3}
\mathcal{Q} = \left\{ {{\rho _1},...,{\rho _p},...,{\rho _{{2^q}}}} \right\},
\end{equation}
where $\rho_p$ denotes the $p$th element in the discrete phase shift level set $\mathcal{Q}$, and ${\rho _p} = {\rho _1} + \left( {p - 1} \right) \times \Omega, \ p=1,2,...,2^q.$ 

For the phase quantization transforming the continuous phase shift matrix ${\bf{\Phi }}$ to the discrete counterpart ${\bf{\Delta }}$, we define a quantizer $Q_{\gamma, \rho}^q: \bf{\Phi} \mapsto \bf{\Delta}$, such that
\begin{equation}
\label{eqv2_1}
\begin{array}{c}
\left. {\begin{array}{*{20}{l}}
{{\rm{Quantization\ threshold:\ }}\gamma  \in [0,2\pi )}\\
{{\rm{Quantization\ interval:\ }}\Omega  = 2\pi /{2^q}}\\
{{\rm{Quantization\ range:\ }}[\gamma ,\gamma  + \Omega )}\\
{{\rm{Quantization\ level:\ }}{\rho _p} = {\rho _1} + (p - 1)\Omega }
\end{array}} \right\} 
\to \Delta_{n,m} = Q_{\gamma, \rho}^q(\phi_{n,m}, \gamma, q, \mathcal{Q}),
\end{array}
\end{equation}
where quantization threshold $\gamma$ divides the continuous phase shift range of $[0,2\pi)$ into $2^q$ discrete ranges. Quantization level $\rho_p$ states the mapping relations between the continuous phase shift and its discrete counterpart. For the simplicity of expression, the calculation procedure of phase quantization in \eqref{eqv2_1} can be rewritten as follows.   
\begin{equation}
\label{eq4}
\begin{array}{c}
\Delta _{n,m}={\rho _p}, \ {\phi _{n,m}} \in [\gamma + (p-1) \times \Omega, \ \gamma + p \times \Omega), 
\forall \ p \in \left\{ {1,2,...,{2^q}} \right\}.
\end{array}
\end{equation}

According to \eqref{eq1}, the received power is directly dominated by the superposition of EM field at Rx, which subject to phase alignment. Following the phase quantization criterion described in \eqref{eq4}, we formulate a joint phase quantization and alignment optimization problem for maximizing the received power as \eqref{eq5}. 
\begin{equation}
\label{eq5}
\begin{array}{l}
\gamma  = \arg \max \left( {\xi ({\bm{\Delta }})} \right), \\
\vspace{-0.3cm}
\quad s.t.\ {\rm{ Eq}}{\rm{.\ \eqref{eq4}}}.
\end{array}
\end{equation}
where $\gamma$ specifies the range of ${\bf{\Phi }}$ to be quantized, which depends on phase quantization strategies. Due to the non-linearity and irreversibility of discrete phase alignment, the above optimization problem is challenging to solve.

\section{Characteristic Analyses of Phase Alignment}

\begin{figure}[htbp]
     \centering
     \subfloat[]{\includegraphics[width = 2.8in]{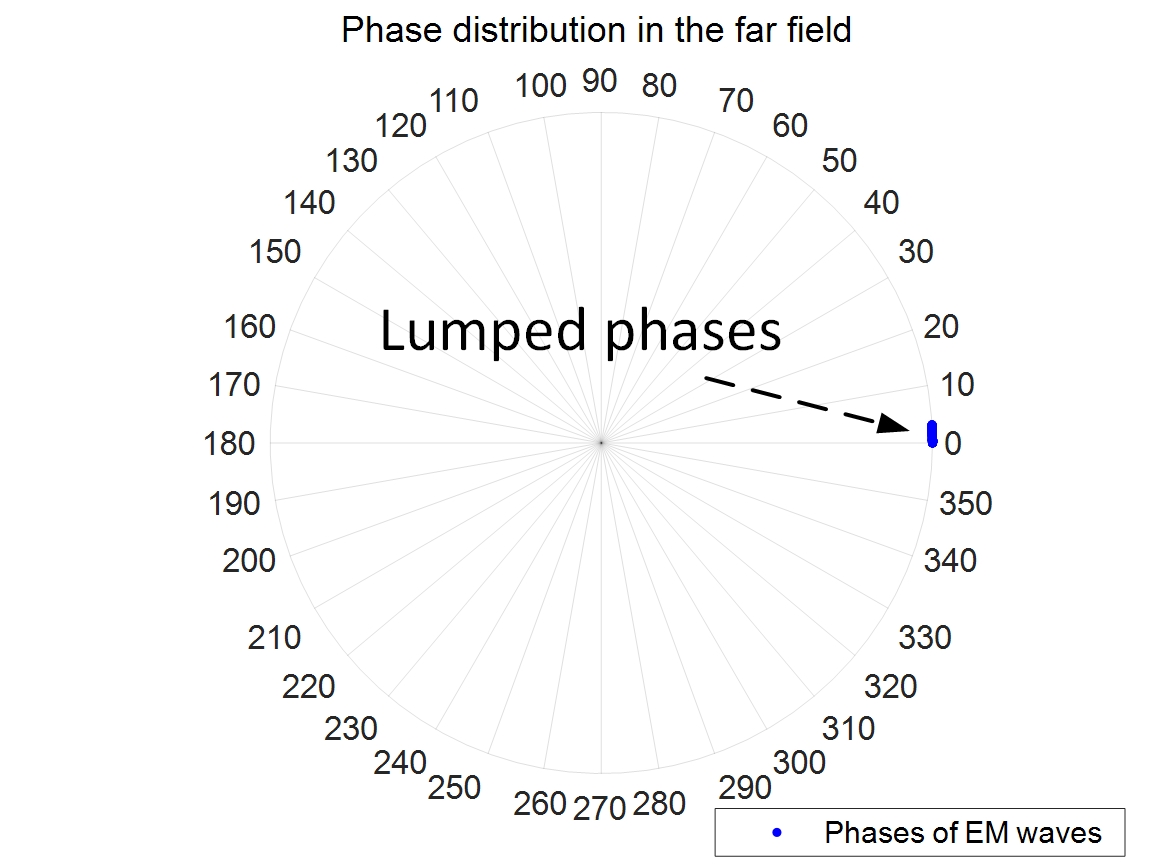}\label{fig_dis_a}}\hspace{1cm}
     \subfloat[]{\includegraphics[width = 2.8in]{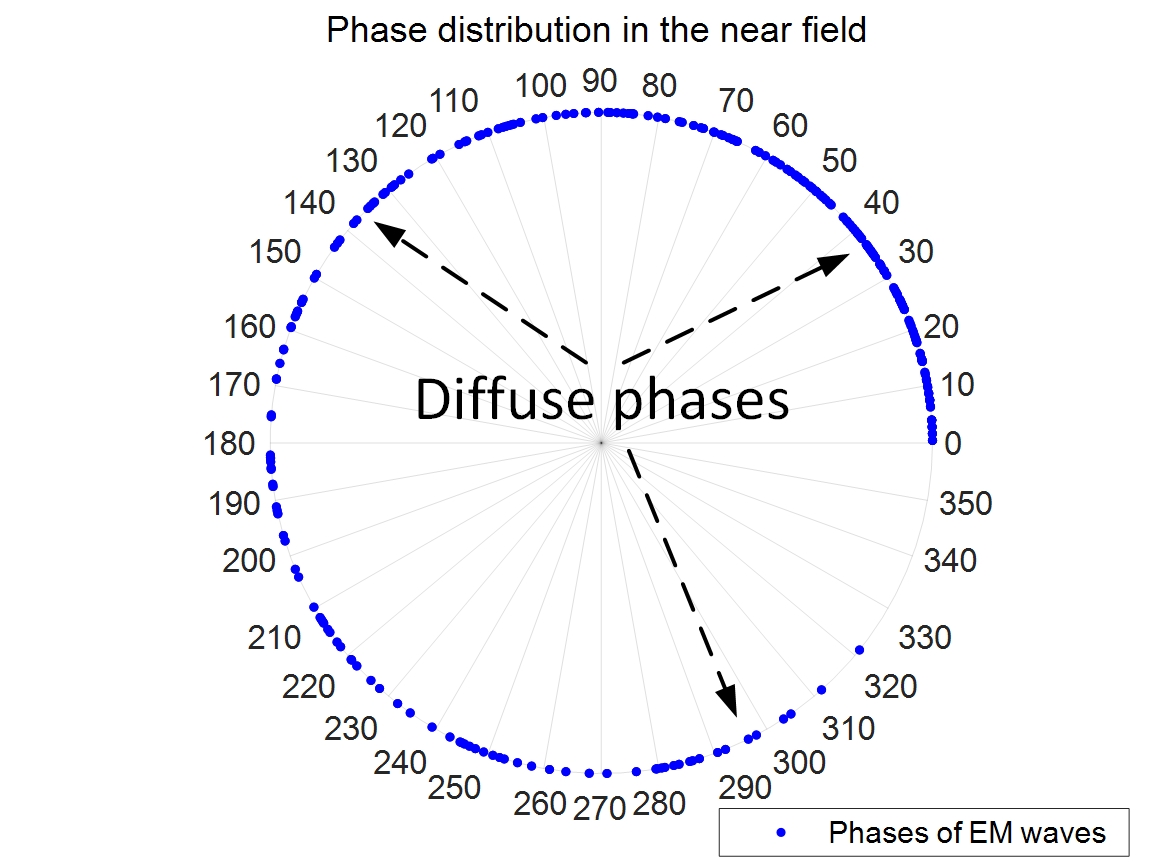}\label{fig_dis_b}} 
    \caption{Schematic diagrams of phase distribution in the far and near fields. (a) In the far field. (b) In the near field. }
\label{fig_dis}
\end{figure}

Taking into account the multi-unit characteristics of RIS, the phase alignment issues could be greatly influenced by the various phase distribution patterns, such as spherical wave front or plane wave front. In this section, we showcase the phase distribution of EM waves and characterize the properties of phase alignment in far field and near field of RIS respectively. It is worth mentioning that in traditional phase quantization method, the quantization threshold $\gamma$ in \eqref{eq5} is set to a constant value \cite{b6}. For example, the discrete phase shift level $\rho$ is usually selected as the fixed quantization threshold $\gamma$, due to the principle of proximity. However, considering the variability on field effect under channel conditions, the traditional fixed-threshold method may perform a large quantization error when the phases are non-uniformly distributed, and further destruct the phase alignment. \reffig{fig_dis} demonstrates the comparison for phase distribution of EM waves in the far field and near field of RIS, respectively. In this figure, the ``blue'' points stand for the phases of EM waves in polar coordinates. The detailed analyses on phase distribution and alignment are provided in the following subsections.

\vspace{-0.5cm}

\vspace{-0.5cm}
\subsection {Far field}
When both Tx and Rx are in the far field of RIS, the wave-path distances of all the EM waves reflected by unit cells are nearly same, considering that the communication distances are much larger than the RIS size. In this case, the plane wave front could be assumed. Then, the ideal continuous phase shifts of all unit cells could be approximatively the same value, according to \eqref{eq_r5}. As shown in \reffig{fig_dis_a}, in the far field case, phase distribution of EM waves is lumped together, which indicate that almost all of the phases are the same values. Then the desired discrete phase shifts of RIS to align phases can be easily determined. As such, for the optimization problem in \eqref{eq5}, the impact of phase quantization threshold on the received power is negligible. In other words, the discrete phase shifts obtained by any quantization threshold are close to the optimum.

\subsection {Near field}
When Tx and/or Rx are in the near field of RIS, the wave-path distances of EM waves reflected by unit cells could be substantially different, taking into account the influence of RIS size. In this case, the spherical wave front could be assumed. Then, the phases of EM waves at Rx could be unaligned and non-uniformly distributed. For the sake of illustration, we provide the example for characterizing the phase distribution in near field, as follows.

\textbf{\textit{Example 1}}: Two unit cells of $\mathrm{U}_{1,1}$ and $\mathrm{U}_{N/2,M/2}$ are taken for example. As depicted in \reffig{fig1}, the wave-path distance of Tx-$\mathrm{U}_{1,1}$-Rx link is 
\begin{equation}
L_{1,1}^{wave} = r_{1,1}^t + r_{1,1}^r, 
\end{equation}
where $r_{1,1}^t$ and $r_{1,1}^r$ can be calculated by \eqref{eq7a} and \eqref{eq7b}, respectively.
\vspace{-0.5cm}
\begin{figure*}[htbp]
\begin{subequations}
\begin{align}
r_{1,1}^t &= \sqrt {{{\left[ {{d_1}\sin ({\theta _t})\cos ({\varphi _t}) - {d_x}\frac{{N - 1}}{2}} \right]}^2} + {{\left[ {{d_1}\sin ({\theta _t})\sin ({\varphi _t}) - {d_y}\frac{{M - 1}}{2}} \right]}^2} + {{\left[ {{d_1}\cos ({\theta _t})} \right]}^2}},  \label{eq7a}\\
r_{1,1}^r &= \sqrt {{{\left[ {{d_2}\sin ({\theta _r})\cos ({\varphi _r}) - {d_x}\frac{{N - 1}}{2}} \right]}^2} + {{\left[ {{d_2}\sin ({\theta _r})\sin ({\varphi _r}) - {d_y}\frac{{M - 1}}{2}} \right]}^2} + {{\left[ {{d_2}\cos ({\theta _r})} \right]}^2}}. \label{eq7b}
\end{align}
\end{subequations}
\vspace{-0.5cm}
\end{figure*}

\vspace{-0.5cm}
The wave-path distance of Tx-$\mathrm{U}_{N/2,M/2}$-Rx link is
\begin{equation}
\label{eq8}
L_{N/2,M/2}^{wave} = r_{N/2,M/2}^t + r_{N/2,M/2}^r, 
\end{equation}
where $r_{N/2,M/2}^t$ and $r_{N/2,M/2}^r$ can be calculated respectively by \eqref{eq_r5a} and \eqref{eq_r5b}. 
\begin{figure*}[!htp]
\begin{subequations}
\begin{align}
r_{N/2,M/2}^t & = \sqrt {{{\left[ {{d_1}\sin ({\theta _t})\cos ({\varphi _t}) - \frac{{d_x}}{2}} \right]}^2} + {{\left[ {{d_1}\sin ({\theta _t})\sin ({\varphi _t}) - \frac{{d_y}}{2}} \right]}^2} + {{\left[ {{d_1}\cos ({\theta _t})} \right]}^2}}, \label{eq_r5a}\\
r_{N/2,M/2}^r & = \sqrt {{{\left[ {{d_2}\sin ({\theta _r})\cos ({\varphi _r}) - \frac{{d_x}}{2}} \right]}^2} + {{\left[ {{d_2}\sin ({\theta _r})\sin ({\varphi _r}) - \frac{{d_y}}{2}} \right]}^2} + {{\left[ {{d_2}\cos ({\theta _r})} \right]}^2}}. \label{eq_r5b}
\end{align}
\end{subequations}
\vspace{-0.5cm}
\end{figure*}

The wave-path difference between the EM waves reflected by these two unit cells can be represented as
\begin{equation}
\label{eq9}
{L_{dif}} = \left| {L_{1,1}^{wave} - L_{N/2,M/2}^{wave}} \right|.
\end{equation}

For illustrative conveniences, consider the propagation parameters in the near field as $\theta_r=\theta_t=\pi/4$, $\varphi_t=0$, $\varphi_r=\pi$, $d_1=d_2=10\lambda$, $N=16$, $M=32$, $d_x=d_y=\lambda/2$, then the wave-path difference in \eqref{eq9} is calculated to be
\begin{equation}
\label{eq10}
{L_{dif}} \approx 6.07\lambda  \gg \lambda .
\end{equation}

{\textit{Example 1}} illustrates the phenomenon that the EM waves reflected by unit cells may have large wave-path differences among one another when Tx and/or Rx in the near field of RIS. This could result in the diffuse and non-uniform phase distribution of EM waves at Rx, as exhibited in \reffig{fig_dis_b}, which further destructs the phase alignment. Under such circumstances, the appropriate discrete phase shifts configured by RIS need to be elaborately designed to align phases, which are directly dominated by phase quantization strategies. However, as mentioned above, the traditional quantization method with a constant quantization threshold may cause quantization errors in this case. In consequence, the innovative phase quantization strategies with flexible and adjustable quantization thresholds are pre-requisite to be designed and regularized for strengthening beamforming of RIS. 


\section{The Proposed Theorem and Methods}
\label{sec3}

To enhance beamforming provided by RIS, the phases of EM waves at Rx should be aligned through tuning discrete phase shifts of unit cells. However, as mentioned above, the phase distribution could be non-uniform and unaligned, which challenges the performance of traditional phase quantization method. In this section, for the purpose of maximizing beamforming effect, the underlying phase distribution laws are revealed and proved firstly. Then its DoF of optimal solution is analyzed. Note that the DoF herein refers to the capacity of optimal solution space, i.e., how many phase distribution pattern satisfying the law. In addition, it is explained that these solutions can be determined accordingly by setting the appropriate phase quantization thresholds. On this basis, two phase quantization methods aiming at maximizing phase alignment in \eqref{eq5}, i.e., DTPQ and EIPQ, are proposed for transforming ideal continuous phase shifts into practical discrete counterparts for RIS-assisted channel.

{\bf{Theorem 1:}} Let $\phi _{j,i}^q$ and $\phi _{n,m}^q$ denote the quantized phases of EM waves for Tx-$\mathrm{U}_{j,i}$-Rx and Tx-$\mathrm{U}_{n,m}$-Rx, respectively. For the optimal solution to \eqref{eq5}, we have 
\begin{equation}
\label{eq_r6}
\begin{array}{l}
\max \left( {\left| {\phi _{j,i}^q - \phi _{n,m}^q} \right|} \right) \le \Omega,\\
\vspace{-0.3cm}
\quad \forall \ i, m \in \left\{ {1,2,...,{M}} \right\},\\
\quad \forall \ j, n \in \left\{ {1,2,...,{N}} \right\}.
\end{array}
\end{equation}

{\bf{Proof:}} See Appendix.

{\textbf{Theorem 1}} says that, to ensure phase alignment in discrete space, any two quantized phases should be within a quantization interval. It illustrates the characteristics of phase quantization from the concept of the quantized phase distribution laws, which provides a perspective for designing quantization strategies. Note that as $q$ tends to infinity, discrete phase alignment boils down to perfect continuous phase alignment. From the phase distribution laws described in  {\textbf{Theorem 1}}, we can infer that for a RIS with $MN$ unit cells, there are $MN$ kinds of phase distribution patterns to satisfy \eqref{eq_r6}, which are explained in detail as {\textit{Example 2}}. Note that for the sake of illustration, we consider the 1-bit phase resolution with discrete phase shift levels of $0$ and $\pi$ in {\textit{Example 2}}, whose conclusion still holds when the discrete phase shift levels change or when it is extended to multi-bit phase resolution. 

\textbf{\textit{Example 2}}: Denote the phase distribution matrix at Rx after phase quantization as ${\bf{\Phi^Q}}$. In this example, we analyze all of the possible phase distribution patterns for a RIS composed of $MN$ unit cells with {\textbf{Theorem 1}} satisfied. They are orderly displayed one by one, as follows. \\ 
Phase distribution pattern $1$: ${\bf{\Phi^Q}} \in [{\phi _{1,1}},{\phi _{1,1}} +\pi)$, which can be identified by $\gamma=\phi_{1,1}$: 
\begin{equation}
\begin{array}{l}
\phi _{n,m}^q = \left\{ {\begin{array}{*{20}{l}}
{{\phi _{n,m}}, \ {\phi _{n,m}} \in [{\phi _{1,1}},{\phi _{1,1}} + \pi )},\\
{{\phi _{n,m}}+ \pi, \ {\phi _{n,m}} \in [{\phi _{1,1}} + \pi ,{\phi _{1,1}} + 2\pi )},
\end{array}} \right. \forall m \in \left\{ {1,2,...,{M}} \right\}, \forall n \in \left\{ {1,2,...,{N}} \right\}.
\end{array}
\nonumber
\end{equation}
Phase distribution pattern $2$: ${\bf{\Phi^Q}} \in [{\phi _{1,2}},{\phi _{1,2}} +\pi)$, which can be identified by $\gamma=\phi_{1,2}$: 
\begin{equation}
\begin{array}{l}
\phi_{n,m}^q = \left\{ {\begin{array}{*{20}{l}}
{{\phi _{n,m}}, \ {\phi _{n,m}} \in [{\phi _{1,2}},{\phi _{1,2}} + \pi )},\\
{{\phi _{n,m}}+\pi, \ {\phi _{n,m}} \in [{\phi _{1,2}} + \pi ,{\phi _{1,2}} + 2\pi )},
\end{array}} \right. \forall m \in \left\{ {1,2,...,{M}} \right\}, \forall n \in \left\{ {1,2,...,{N}} \right\}.
\end{array}
\nonumber
\end{equation}
\qquad \qquad \qquad \qquad \qquad \qquad \qquad \qquad \qquad \qquad $\bm{\vdots}$ \\
Phase distribution pattern $MN$: ${\bf{\Phi^Q}} \in [{\phi_{N,M}},{\phi_{N,M}} +\pi)$, which can be identified by $\gamma=\phi_{N,M}$:
\begin{equation}
\begin{array}{l}
\phi _{n,m}^q = \left\{ {\begin{array}{*{20}{l}}
{{\phi _{n,m}}, \ {\phi _{n,m}} \in [{\phi _{N,M}},{\phi _{N,M}} + \pi )},\\
{{\phi _{n,m}}+ \pi, \ {\phi _{n,m}} \in [{\phi _{N,M}} + \pi ,{\phi _{N,M}} + 2\pi )},
\end{array}} \right. \forall m \in \left\{ {1,2,...,{M}} \right\},
\forall n \in \left\{ {1,2,...,{N}} \right\}.
\end{array}
\nonumber
\end{equation}

\textit{Example 2} exemplify that the number of possible phase distribution patterns satisfying Theorem 1 equals to the number of unit cells on RIS. The phase distribution patterns therein can be respectively achieved by setting the quantization thresholds equivalent to the elements in ${\bf{\Phi }}$. As a result, it can be inferred that the DoF of optimal solution to \eqref{eq5} is $MN$ in the worst case, rather than the well-known $2^{qMN}$ in exhaustive search method. It is also worth mentioning that the DoF of optimal solution is only related to the number of unit cells and irrelevant to the number of bits, according to \textit{Example 2}. By reason of the foregoing, we can perform ergodic search strategy in optimal solution space to obtain the desired quantization threshold, with the complexity of $O(MN)$. Then we can solve the joint optimization problem described in \eqref{eq5}. Following this guideline, in next subsections, the DTPQ method with \textbf{optimality} and a linear complexity is proposed. Moreover, we also propose a EIPQ method with sub-optimality yet with an even lower complexity. 

\begin{table}[htbp]
\label{algorithm1}
    \centering
     \begin{threeparttable}[b]
\begin{tabular}{rl}
\toprule
\multicolumn{2}{l}{\textbf{Algorithm 1} DTPQ method for \eqref{eq5}.} \\ 
\midrule
1       & Acquire the information on RIS-assisted SISO system including the locations of Tx, Rx and RIS.  \\
2       & Calculate the ideal continuous phase shift matrix ${\bf{\Phi }}$ as \eqref{eq_r5}.  \\
3       & Generate the quantization threshold matrix ${\bf{\Gamma }}$ according to \eqref{v1_eq1}.  \\
4       & Calculate the desired quantization threshold $\tilde \gamma \in {\bf{\Gamma }}$ as \textbf{Step 5-11}.\\
5       & Initialization $\tilde \xi$=0. \\
6       & \textbf{For} $x=1:N$ \\
7       & \quad \textbf{For} $y=1:M$ \\
8       & \quad \quad Set $\gamma_{x,y} \in {\bf{\Gamma }}$ as the quantization threshold. \\ 
9       & \quad \quad Calculate the discrete phase shifts matrix ${\bf{\Delta }}$ under quantization threshold $\gamma_{x,y}$ according to \eqref{eq4}. \\
10      & \quad \quad Calculate the EM fields $\xi_{x,y}$ under quantization threshold $\gamma_{x,y}$ according to \eqref{eq2}. \\
11      & \quad \quad \textbf{If} ${\xi _{x,y}} > \tilde \xi$ : \textbf{do} $\tilde \xi = {\xi _{x,y}} $, $\tilde \gamma = {\gamma _{x,y}} $. \\
        & \quad \quad \textbf{Else} : no-operation. \\
        & \quad \quad \textbf{End} \\
        & \quad \textbf{End} \\
        & \textbf{End} \\
12      & Output $\tilde \gamma$ as the desired quantization threshold. \\
\bottomrule
\end{tabular}
   \end{threeparttable}
\end{table}

\subsection{DTPQ method}
In this subsection, based on \textbf{Theorem 1} and the analyses on DoF, a DTPQ method is introduced to solve the joint optimization problem described in \eqref{eq5}, by conducting ergodic search in the optimal solution space. In DTPQ, the quantization threshold matrix ${\bf{\Gamma }}$ can be generated as follows.

\begin{equation}
\label{v1_eq1}
\begin{array}{l}
{\bf{\Gamma }} = \left[ {\begin{array}{*{20}{c}}
{{\gamma _{1,1}}}& \cdots &{{\gamma _{1,m}}}& \cdots &{{\gamma _{1,M}}}\\
 \vdots & \ddots &{}&{}& \vdots \\
{{\gamma _{n,1}}}&{}&{{\gamma _{n,m}}}&{}&{{\gamma _{n,M}}}\\
 \vdots &{}&{}& \ddots & \vdots \\
{{\gamma _{N,1}}}& \cdots &{{\gamma _{N,m}}}& \cdots &{{\gamma _{N,M}}}
\end{array}} \right] \in \mathbb{C}^{M \times N},
\end{array}
\end{equation}
where ${\gamma _{n,m}}$ can be calculated by
\vspace{-0.2cm}
\begin{equation}
{\gamma _{n,m}} = {\phi _{n,m}}, \forall \ m \in \left\{ {1,2,...,{M}} \right\}, \forall \ n \in \left\{ {1,2,...,{N}} \right\}.
\end{equation}
Then, an ergodic search of ${\bf{\Gamma}}$ is performed to determine the desired phase quantization threshold. The detailed algorithm procedure of DTPQ is described in Algorithm 1.

\begin{figure*}[htpb]
\vspace{-0.5cm}
     \centering
       \subfloat[]{\includegraphics[width = 3.2in]{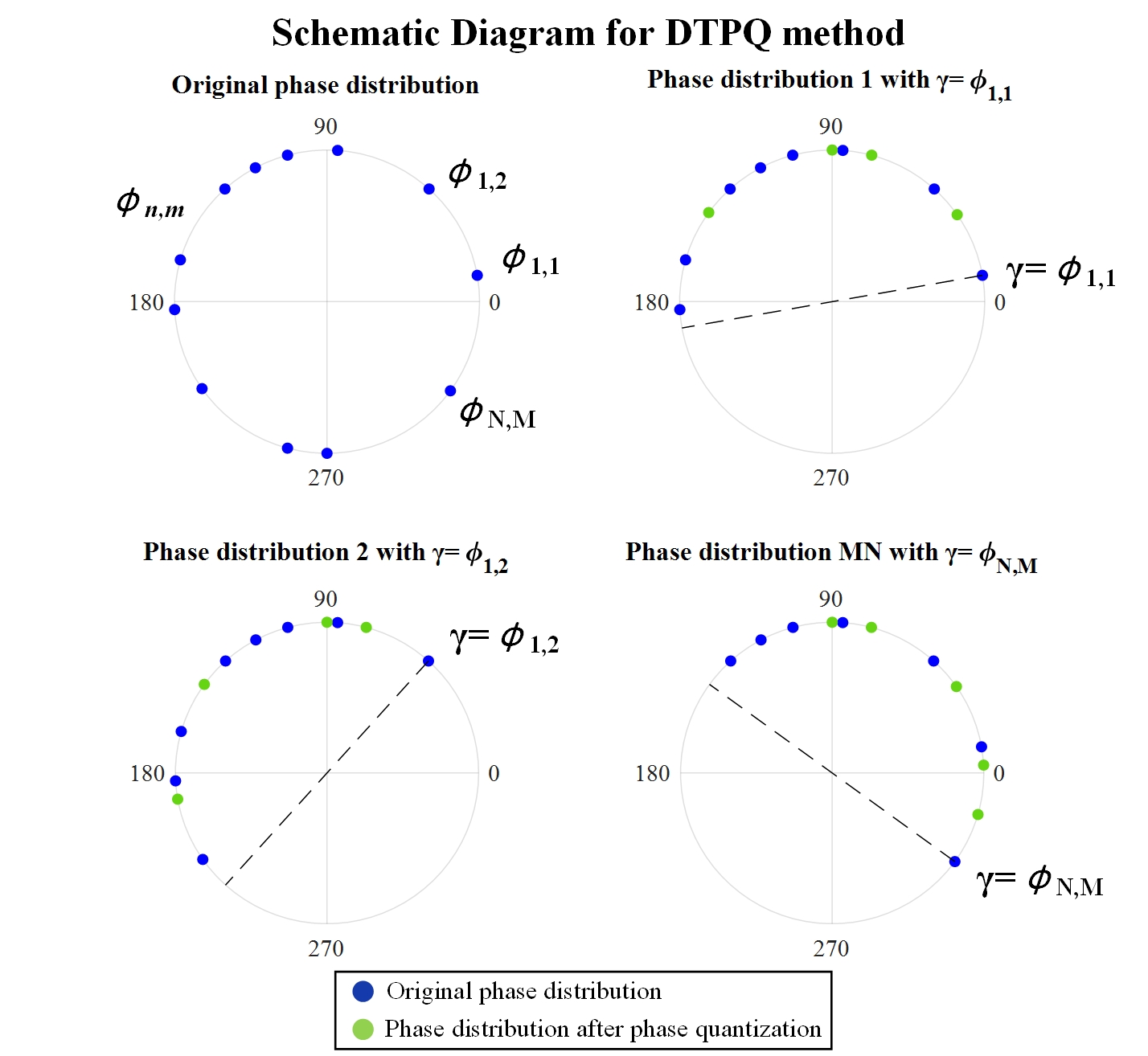}\label{fig 2a}}
       \subfloat[]{\includegraphics[width = 3.2in]{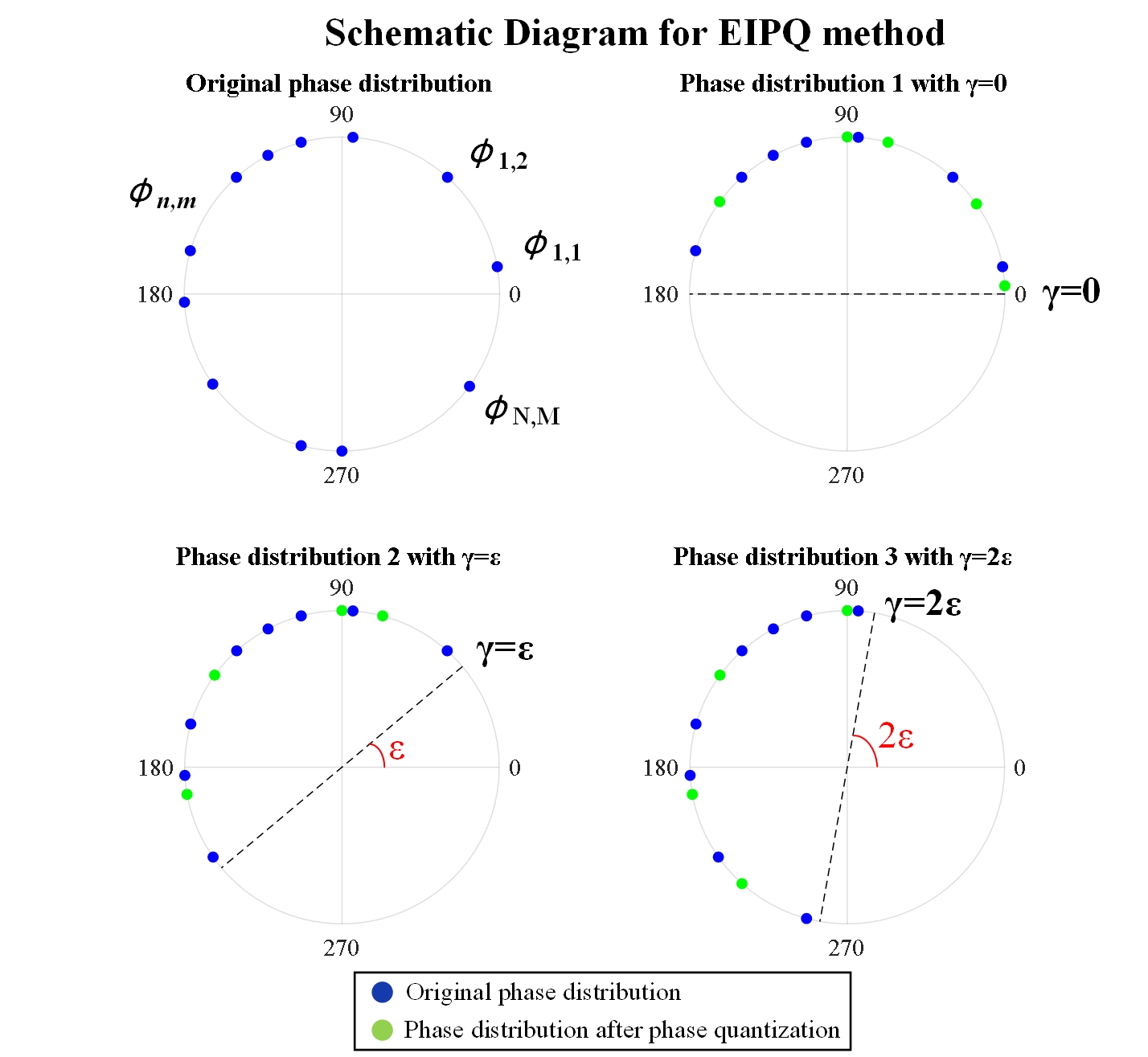}\label{fig 2b}}
    \caption{Schematic diagrams for proposed methods in 1-bit discrete space. (a) DTPQ. (b) EIPQ. }
\label{fig2}
\vspace{-0.5cm}
\end{figure*}

For illustrative conveniences, \reffig{fig 2a} demonstrates the schematic diagrams of the proposed DTPQ method in 1-bit discrete space. In this figure, the ``blue points'' stand for the original phase distribution of EM waves without phase shifts of RIS and the ``green points'' represent the quantified phase distribution with phase shifts of RIS after phase quantization. The ``dotted line'' denotes the quantization threshold $\gamma$ designed by DTPQ, which has a quantization range of [$\gamma$, $\gamma+\pi$) in 1-bit discrete space according to \eqref{eq4}. As can be seen from \reffig{fig 2a}, the quantization thresholds in DTPQ method are determined respectively by the elements in ${\bf{\Gamma }}$, as described in \eqref{v1_eq1}. By searching these phase quantization thresholds in turn, their corresponding discrete phase shift matrix can be calculated. That is, $MN$ kinds of possible solutions to \eqref{eq5} are designed by DTPQ. Then, the discrete phase shift matrix and its phase quantization threshold having the maximum magnitude of EM field can be regarded to be optimal solution to \eqref{eq5}. As a result, the complexity of DTPQ method is $O(MN)$, which is significantly reduced from $O({2^{qMN}})$ in exhaustive search method, while guaranteeing the \textbf{optimality}.

\subsection{EIPQ method}

As a sub-optimal alternative to DTPQ yet with an even lower complexity, we propose EIPQ method to calculate the desired solution to \eqref{eq5} in this subsection. In EIPQ method, the quantization threshold set ${\cal E}$ are generated with an equal interval $\varepsilon$ in [0, $2\pi /{2^q}$) in $q$-bit discrete space, as \eqref{v1_eq2}. 
\begin{equation}
\label{v1_eq2}
\begin{array}{c}
{\cal E} = \{ {\gamma _1}, ... , {\gamma _k}, ... ,{\gamma _K}\} ,
\end{array}
\end{equation}
where $\gamma_k$ can be calculated by
\begin{equation}
\begin{array}{c}
\gamma_k=(k - 1)\varepsilon, \forall \ k \in \left\{ {1,2,...,{K}} \right\}. 
\end{array}
\end{equation}
$K$ is the number of elements in ${\mathcal{E}}$, which can be defined by $(K-1)\varepsilon  < 2\pi /{2^q} \le K\varepsilon$. $\varepsilon$ represents the quantization step, which is preset artificially. Through an ergodic search in ${\mathcal{E}}$, we could find the desired quantization threshold and its corresponding discrete phase shifts for \eqref{eq5}. The detailed algorithm procedure of EIPQ is described in Algorithm 2. 

\begin{table}[htbp]
\vspace{-0.5cm}
\label{algorithm2}
    \centering
     \begin{threeparttable}[b]
\begin{tabular}{rl}
\toprule
\multicolumn{2}{l}{\textbf{Algorithm 2} EIPQ method for \eqref{eq5}.} \\ 
\midrule
1       & Acquire the information on RIS-assisted SISO system including the locations of Tx, Rx and RIS.  \\
2       & Calculate the ideal continuous phase shift matrix ${\bf{\Phi }}$ as \eqref{eq_r5}.  \\
3       & Generate the quantization threshold set ${\mathcal{E}}$ according to \eqref{v1_eq2}.  \\
4       & Calculate the desired quantization threshold $\tilde \gamma \in {\mathcal{E}}$ as \textbf{Step 5-10}.\\
5       & Initialization $\tilde \xi$=0. \\
6       & \textbf{For} $x=1:K$ \\
7       & \quad Set $\gamma_{x} \in {\mathcal{E}}$ as the quantization threshold. \\ 
8       & \quad Calculate the discrete phase shifts matrix ${\bf{\Delta }}$ under quantization threshold $\gamma_{x}$ according to \eqref{eq4}. \\
9      & \quad Calculate the EM fields $\xi_{x}$ under quantization threshold $\gamma_{x}$ according to \eqref{eq2}. \\
10      & \quad \textbf{If} ${\xi _{x}} > \tilde \xi$ : \textbf{do} $\tilde \xi = {\xi _{x}} $, $\tilde \gamma = {\gamma _{x}} $. \\
        & \quad \textbf{Else} : no-operation. \\
        & \quad \textbf{End} \\
        & \textbf{End} \\
11      & Output $\tilde \gamma$ as the desired quantization threshold. \\
\bottomrule
\end{tabular}
   \end{threeparttable}
   \vspace{-0.5cm}
\end{table}

\reffig{fig 2b} exhibits the schematic diagram of EIPQ in 1-bit discrete space. In this figure, the ``blue points'' stand for the original phase distribution of EM waves without phase shifts of RIS and the ``green points'' represent the quantified phase distribution with phase shifts of RIS after phase quantization. The ``dotted line'' denotes the quantization threshold $\gamma$ determined by EIPQ, which has a quantization range of [$\gamma$, $\gamma+\pi$) in 1-bit discrete space. As can be seen from this figure, the quantization thresholds in EIPQ are determined with an equal step $\varepsilon$ in sequence. Then, $K$ possible solutions to \eqref{eq5} are designed by EIPQ. The quantization threshold and its corresponding discrete phase shift matrix under the maximum magnitude of EM field are considered to be desired ones. In consequence, the complexity of EIPQ is $O(K)$ and $K = \left \lfloor 2\pi /({2^q}\varepsilon) \right \rfloor \ll MN$, where $\left \lfloor \bullet \right \rfloor$ refers to the floor function. Note that the complexity of EIPQ is only related to the preset step $\varepsilon$ and the number of bits $q$. Thus, it can be regarded as constant considering the fact that $q$ and $\varepsilon$ are usually invariables. The complexity of EIPQ is further reduced from DTPQ, though EIPQ comes at the price of a slight decrease in beamforming performance compared to DTPQ yet still performs better than traditional quantization method. 

The comparisons on performance, complexity, and influence factors of the proposed DTPQ method, the proposed EIPQ method, the exhaustive search method and the traditional fixed-threshold method are summarized in \autoref{table_r1}. It is worth mentioning that the performance achieved by the proposed DTPQ method can be regarded as the upper bound in discrete space, due to its \textbf{optimality} (equivalent to exhaustive search method) for aligning phases. In addition, note that except for the communication scenario considered in this paper, the proposed DTPQ and EIPQ methods are also suitable for more general RIS-assisted SISO scenarios, such as: 1) The RIS with signal amplification capability is deployed to assist communication; 2) Both of the Tx-Rx direct link and the Tx-RIS-Rx cascaded link are considered in RIS-assisted communication system; 3) Tx and/or Rx are located in either the far field or the near field of RIS.

\begin{table*}[htbp]
\caption{Comparison of phase quantization methods}
\label{table_r1}
    \centering
\begin{tabular}{|c|c|c|c|c|}
\hline
\textbf{Method}  & Exhaustive search  & DTPQ  & EIPQ  & Traditional fixed-threshold
\\ \hline
\textbf{Performance}  & Optimal  & Optimal 
& Sub-optimal     & Low-performing      
\\ \hline
\textbf{Complexity}  & Exponential ($O(2^{qMN})$)  & Linear ($O(MN)$)  & Constant ($O(K)$)  & Constant ($O(1)$)
\\ \hline
\multirow{2}{*}{\tabincell{c}{\textbf{Influence factor}}} & \multirow{2}{*}{\tabincell{c}{Number of bits, \\ Number of unit cells}} & \multirow{2}{*}{\tabincell{c}{Number of unit cells}} & \multirow{2}{*}{\tabincell{c}{Number of bits, \\ preset step}} & \multirow{2}{*}{\tabincell{c}{None}} \\
& & & & 
\\ \hline
\end{tabular}
\end{table*}

\section{Simulation Results and Field Trials}
\label{sec4}

In this section, we demonstrate the simulation results of the proposed methods against the traditional fixed-threshold quantization method, covering beamforming performance, complexity as well as robustness, in 1-bit and 2-bit discrete spaces respectively. It is worth mentioning that the traditional fixed-threshold method adopts the constant value as phase quantization threshold, which is equivalent to the phase shift level of the applied RIS. Moreover, considering the \textbf{optimality} of DTPQ, we analyze the PL scaling laws in discrete space under phase shifts designed by DTPQ, which are also compared with those in continuous space. In addition, two field trials are carried out at 2.6 GHz and 35 GHz frequency bands respectively, to validate the performance of the proposed method in practical environment.

\subsection{Simulation results}

The simulation performance of the proposed methods are exhibited and compared with that of traditional fixed-threshold method in this subsection. The communication scenario shown in \reffig{fig1} is considered. Two RISs operating at the current commercial communication frequency bands of 2.6 GHz and 4.9 GHz respectively are taken into account \cite{v1_b1}, which concern a variety of bit resolutions. The detailed parameter information on these two RISs and the simulation settings are summarized in \autoref{table2}. Unless otherwise specified, all parameters are set to the values in \autoref{table2} by default, whereas the settings in each figure take precedence wherever applicable.

\begin{table}[t]
\vspace{-0.2cm}
\caption{Simulation parameters}
\label{table2}
    \centering
\begin{tabular}{|c|c|c|}
\hline
\textbf{Parameter}              & \textbf{RIS 1}     & \textbf{RIS 2}       \\ \hline
Operating frequency $f$         & 2.6 GHz            & 4.9 GHz              \\ \hline
Wavelength $\lambda$            & $\lambda_1$=0.115 m     & $\lambda_2$ = 0.061 m              \\ \hline
Number of rows $M$              & 32                 & 50                   \\ \hline
Number of columns $N$           & 16                 & 25                   \\ \hline
Number of unit cells      & 32 $\times$ 16 = 512   & 50 $\times$ 25 = 1250     \\ \hline
Size of unit cell               & $\lambda_1 /2 \times \lambda_1 /2$  & $\lambda_2 /2 \times \lambda_2 /2$  \\ \hline
Phase resolution $q$  & 1-bit  & 2-bit    \\ \hline
Discrete phase shift levels & $\rho_1$=55°, $\rho_2$=235° & $\rho_1$=0°, $\rho_2$=90°, $\rho_3$=180°, $\rho_4$=270° \\ \hline
Tx-RIS distance $d_1$           & 10 m      & 10 m      \\ \hline
Rx-RIS distance $d_2$           & 5:0.1:10 m         & 50:0.1:55 m          \\ \hline
EoA $\theta_t$ & 45°            & 45°                  \\ \hline
EoD $\theta_r$ & 45°            & 45°                  \\ \hline
AoA $\varphi_t$  & 0°           & 0°                   \\ \hline
AoD $\varphi_r$  & 180°         & 180°                 \\ \hline
Antenna gain of Tx $G_t$   &   8.25 dBi   & 8.25 dBi   \\ \hline
Antenna gain of Rx $G_r$   &   8.25 dBi   & 8.25 dBi   \\ \hline
Transmitted power $P_t$         & 0 dBm              & 0 dBm                \\ \hline
\end{tabular}
\vspace{-0.5cm}
\end{table}

\begin{figure}[htbp]
\vspace{-0.5cm}
\centering
\begin{minipage}[t]{0.48\textwidth}
\centering
\includegraphics[width=3in]{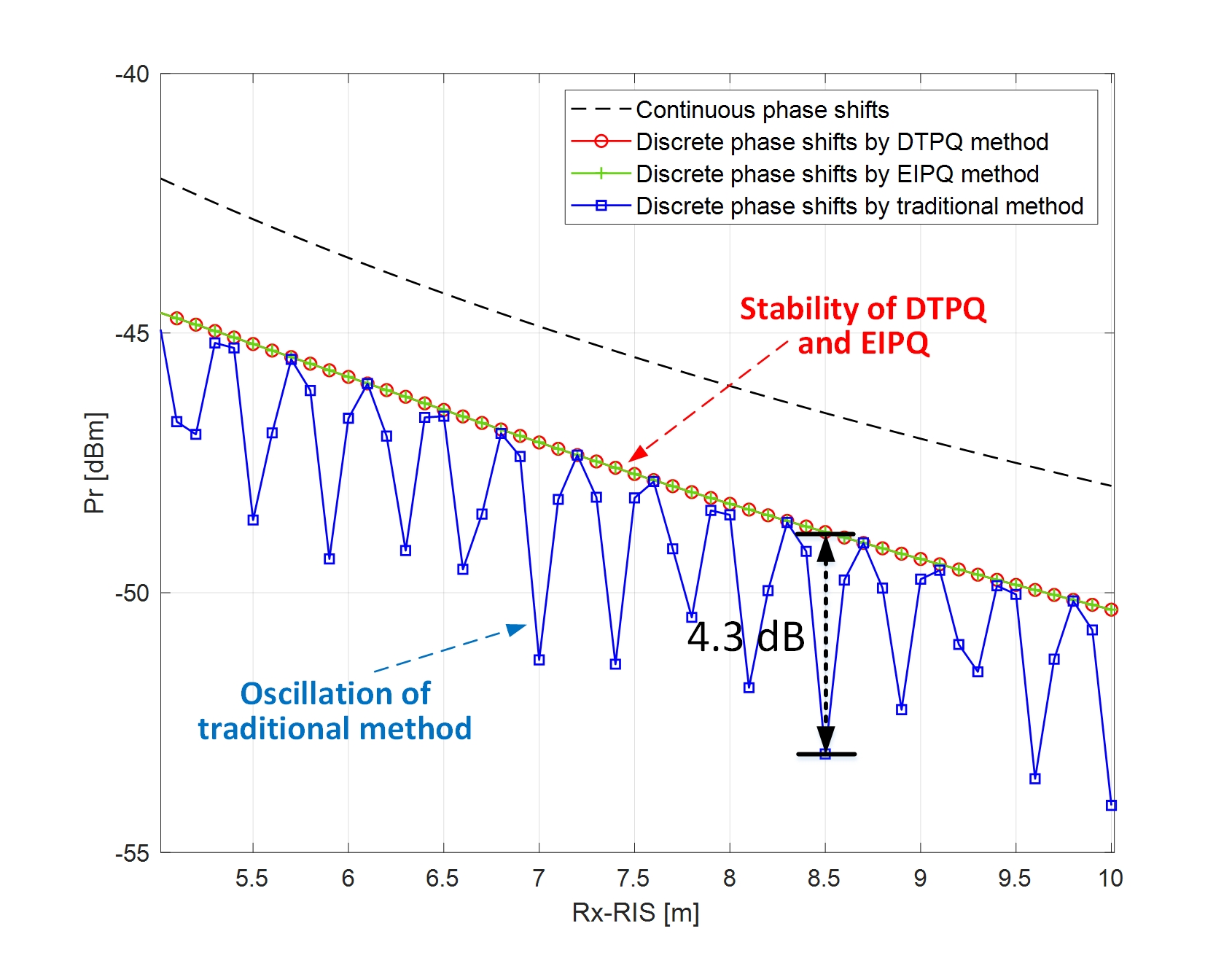}
\vspace{-0.5cm}
\caption{Received power v.s. Rx-RIS distance.}
\label{fig3}
\end{minipage}
\begin{minipage}[t]{0.48\textwidth}
\centering
\includegraphics[width=3in]{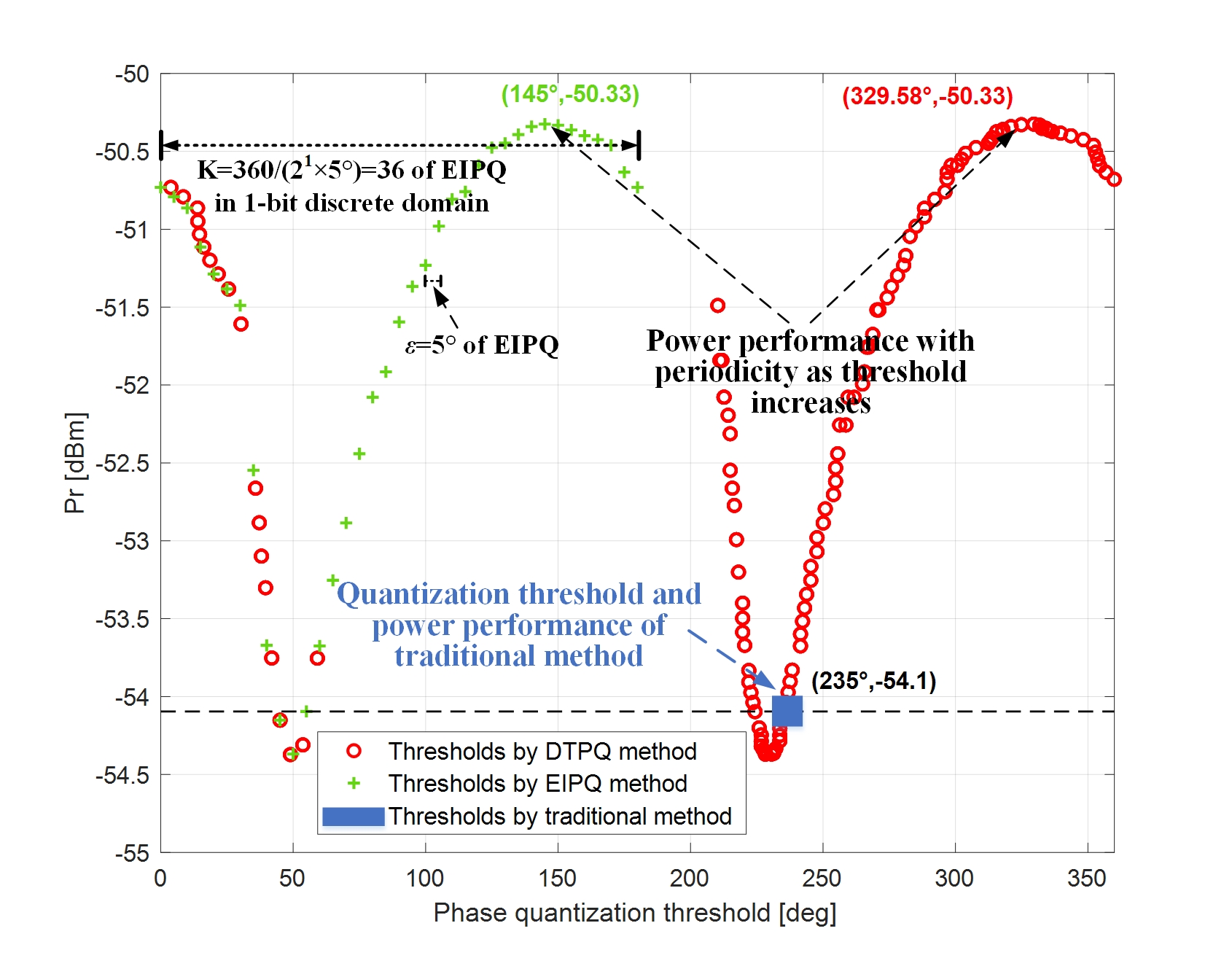}
\vspace{-0.5cm}
\caption{Received power v.s. Phase quantization threshold.}
\label{fig4}
\end{minipage}
\vspace{-0.5cm}
\end{figure}

\reffig{fig3} demonstrates the received power $P_r$ with respect to the Rx-RIS distance $d_2$ in 1-bit discrete space, where the \textbf{RIS 1} in \autoref{table2} is taken into account. In this figure, the ``black dotted line'' refers to the performance achieved by configuring continuous phase shifts $\bf{\Phi}$ of RIS. The ``red line'', ``green line'' and ``blue line'' denote the performances achieved by discrete phase shifts $\bf{\Delta}$ respectively designed by DTPQ method, by EIPQ method and by traditional fixed-threshold method. The traditional method adopts 235° (i.e., $\rho_2$ of \textbf{RIS 1} in \autoref{table2}) as the fixed quantization threshold. Note that for EIPQ method in this figure, $\varepsilon$ is set to 5° and $K$=360°/($2^1\times$5°)=36. As can be seen from \reffig{fig3}, the traditional method shows ``oscillation'' phenomena as distance increases, resulting from that the non-uniform and unaligned phase distribution in the near field, which is consistent with the previous analyses. Meanwhile, our proposed DTPQ and EIPQ methods show remarkable stability as well as significant improvements, i.e., the maximum power gain of about 4.3 dB herein, against the traditional fixed-threshold method. It is worth mentioning that in this figure, EIPQ performs almost the same as DTPQ, which indicates its excellent quantization property under its step $\varepsilon$ = 5° while maintaining the low complexity of $O(36)$.

\reffig{fig4} shows the received power $P_r$ with respect to phase quantization thresholds designed by DTPQ method, by EIPQ method and by traditional fixed-threshold method respectively in 1-bit discrete space. In this figure, the Rx-RIS distance is set to 10 m and the other settings are configured as \textbf{RIS 1} in \autoref{table2}. As we can see, the number of quantization thresholds designed by DTPQ is $M \times N=512$, which equals to the number of unit cells on RIS. The number of quantization thresholds designed by EIPQ is $k$=360°/($2^1\times$5°)=36, under its step $\varepsilon$ = 5°. The number of quantization threshold designed by traditional method is 1, due to the fact that its threshold remains 235° all along, i.e., equivalent to the phase shift level $\rho_2$ of \textbf{RIS 1} in \autoref{table2}. In addition, it can be observed that the received power shows periodical variation as quantization threshold increases, where the maximum received powers of -50.33 dBm are achieved at 145° by EIPQ method and at 329.58° by DTPQ method respectively. This phenomenon indicates the periodicity of quantization threshold with the period of $2\pi/2^q$ in discrete space. By contrast, the traditional method only achieves a received power of -54.1 dBm at 235°.

\begin{figure}[htbp]
\vspace{-0.5cm}
\centering
\begin{minipage}[t]{0.48\textwidth}
\centering
\includegraphics[width=3in]{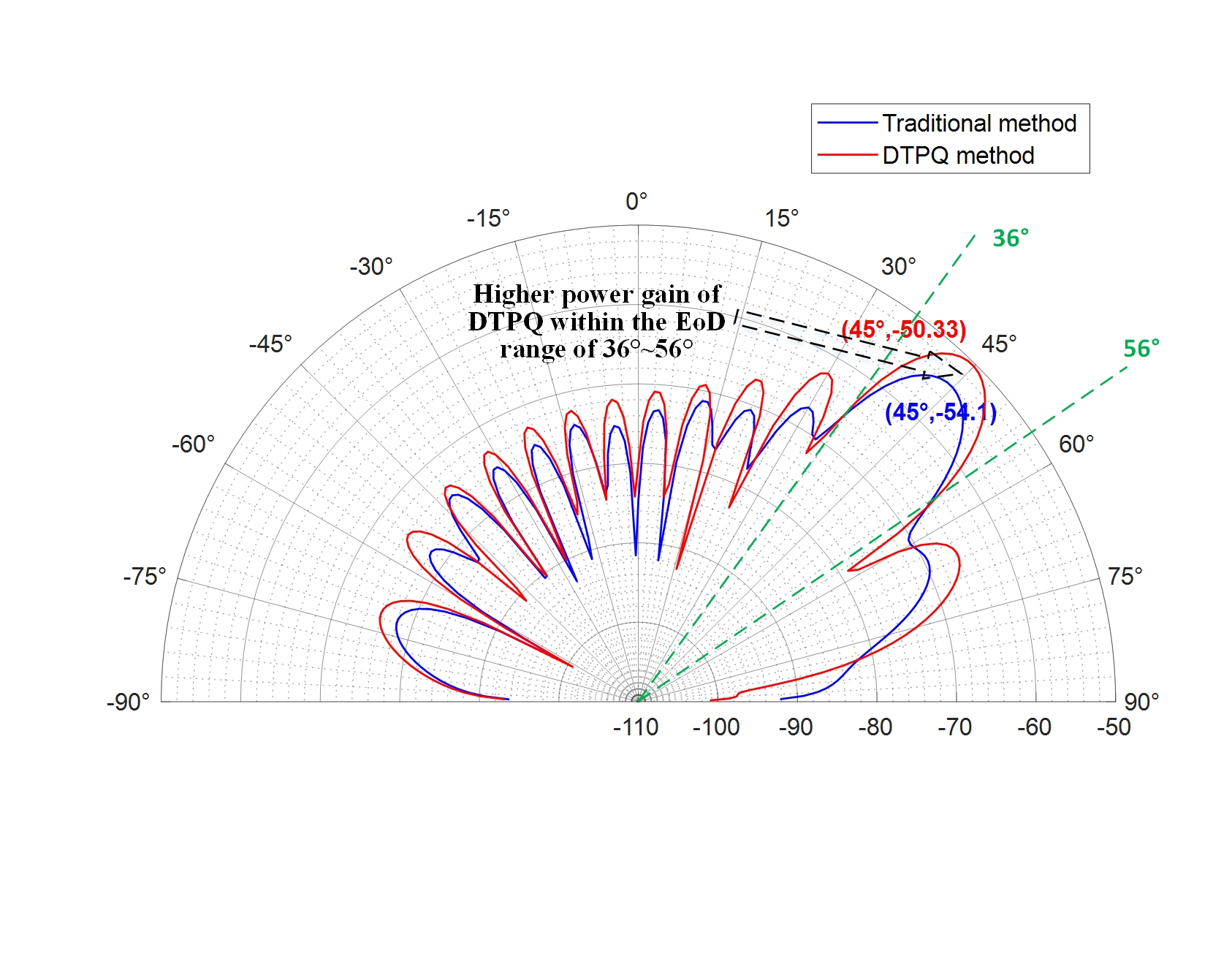}
\vspace{-0.5cm}
\caption{Received power v.s. EoD $\theta_r$.}
\label{fig5}
\end{minipage}
\begin{minipage}[t]{0.48\textwidth}
\centering
\includegraphics[width=3in]{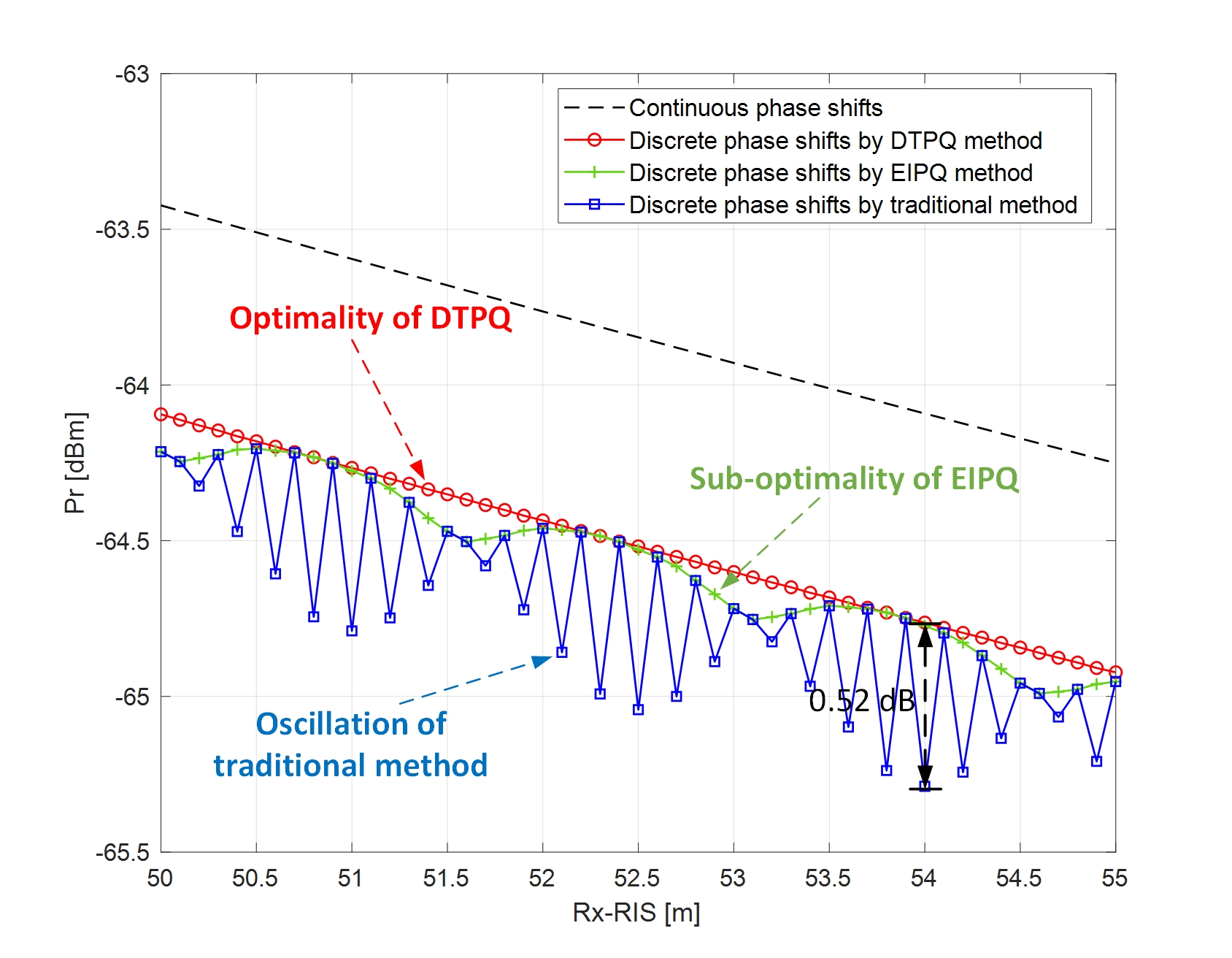}
\vspace{-0.5cm}
\caption{Received power v.s. Rx-RIS distance}
\label{fig_2bit}
\end{minipage}
\vspace{-0.5cm}
\end{figure}

\reffig{fig5} illustrates the received power with EoD $\theta_r$ scanning from -90° to 90°, where the desired EoD is set to 45° and the simulation settings are configured as \textbf{RIS 1} in \autoref{table2}. In this figure, the horizontal axis stands for the received power in the scale of dBm. As can be seen from \reffig{fig5}, at the desired EoD of 45°, the received power of DTPQ method experiences the maximum and shows a prominent improvement, i.e., about 3.77 dB, against that of traditional method. In addition, it is worth mentioning that within the EoD range of 36°$\sim$56°, DTPQ method still performs a higher power gain as compared to the traditional method. This phenomenon indicates that DTPQ has a significant robustness to confront the even slight estimation errors on angles for future application in practical communication environment.

\reffig{fig_2bit} depicts the received power of DTPQ method , EIPQ method, and traditional method respectively, with respect to the Rx-RIS distance $d_2$ in 2-bit discrete space, where the simulation parameters are configured as \textbf{RIS 2} in \autoref{table2}. It should be noted that for EIPQ method in this figure, $\varepsilon$ is set to 45°. The traditional method adopts 270° (i.e., $\rho_4$ of \textbf{RIS 2} in \autoref{table2}) as the fixed quantization threshold. As we can observe, DTPQ method still performs an \textbf{optimality} on power performance, which shows a maximum improvement of about 0.52 dB over the traditional method. Moreover, EIPQ method demonstrates a sub-optimality on power performance, which is slightly worse than DTPQ but still better than traditional method. This results from the fact that the performance of EIPQ degrades gradually when its step $\varepsilon$ becomes larger. In addition, compared to \reffig{fig3}, the ``oscillation'' phenomenon of traditional method in \reffig{fig_2bit} alleviates, originating from the larger number of bits in discrete space. 

\begin{figure*}[htbp]
\vspace{-0.5cm}
     \centering
       \subfloat[]{\includegraphics[height = 2.2in]{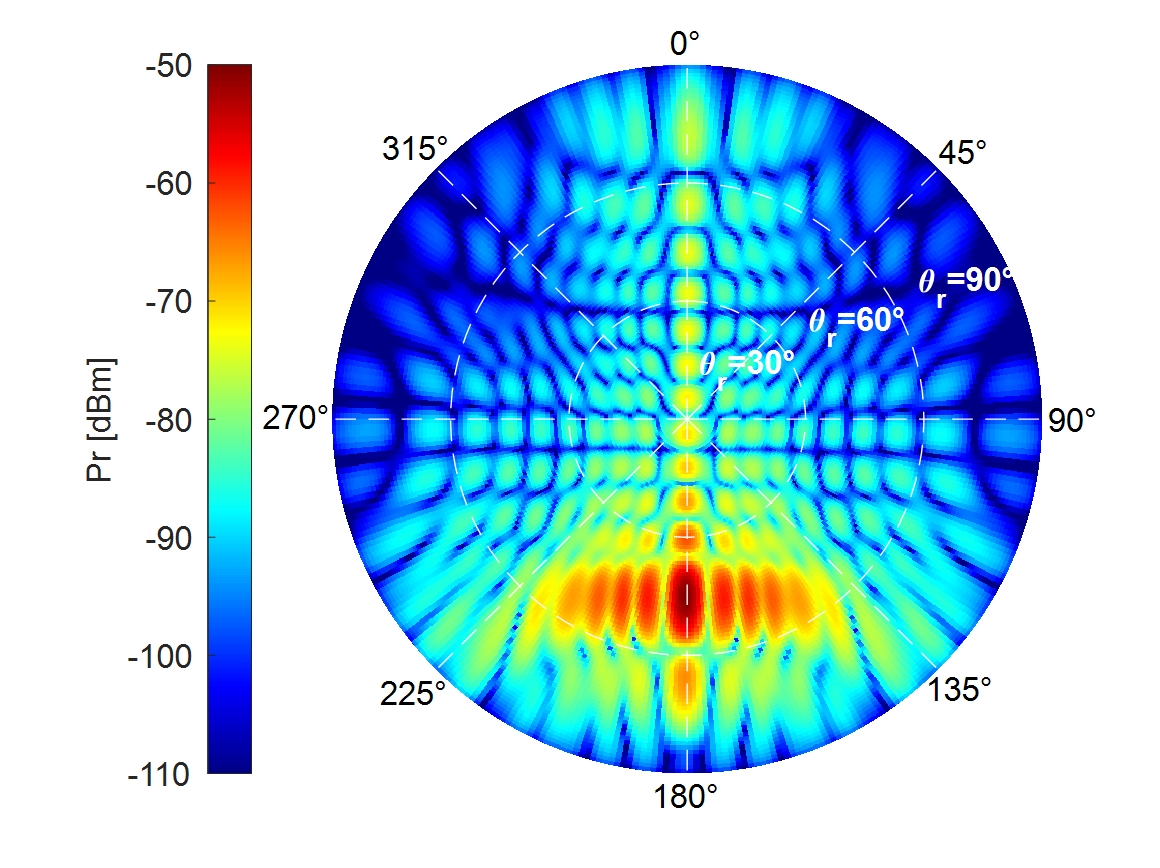}\label{3D_angle_DTPQ}} \hfill
       \subfloat[]{\includegraphics[height = 2.2in]{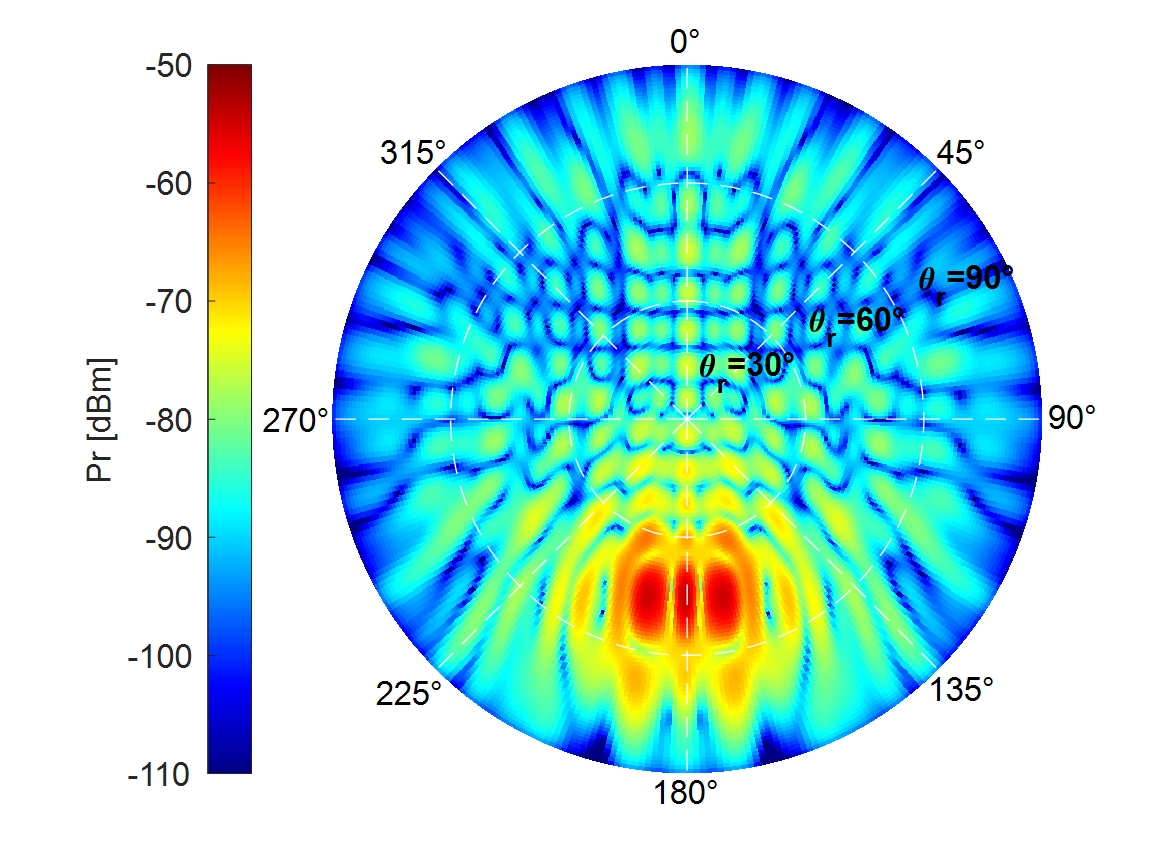}\label{3D_angle_traditional}} 
    \caption{Spatial phase gradient. (a) DTPQ method. (b) Traditional method. }
\label{spatial}
\vspace{-0.5cm}
\end{figure*}

Acting as the crucial role in beamforming, \reffig{spatial} exhibits the shapes of spatial phase gradient with respect to $\varphi_r$ and $\theta_r$, achieved by DTPQ method and traditional method, respectively. In this figure, the desired direction is set to ($\theta_r$=45°, $\varphi_r$=180°), where $d_2$ = 10 m and the other simulation parameters are set as \textbf{RIS 1} in \autoref{table2}. From \reffig{3D_angle_DTPQ}, DTPQ shows its capability of focusing the signal energy significantly at the target angle and produces a stationary spatial phase gradient. On the contrast, the traditional method in \reffig{3D_angle_traditional} shows a diverging signal energy around the target direction, while an irregular spatial phase gradient can be seen throughout the space. 

\subsection{PL scaling laws in discrete space}

As mentioned above, the PL scaling law in continuous space by assuming perfect phase alignment has been reported in many researches, such as \cite{b5}. However, the actual phase shifts being discretized in practical communications, greatly affect the channel characteristics, which may produce differences from the PL scaling law in continuous space. The related research are crucial but still not done. The PL scaling laws in continuous space proposed in \cite{b5} can be expressed as \eqref{eq_r7}.
\begin{equation}
\label{eq_r7}
PL = \frac{{16{\pi ^2}{{({d_1}{d_2})}^2}}}{{{G_t}{G_r}{{(MN{d_x}{d_y})}^2}\cos ({\theta _t})\cos ({\theta _r}){A^2}}}.
\end{equation}
Eq. \eqref{eq_r7} indicates that in continuous space, PL scaling law for RIS-assisted SISO channel is proportional to $(d_1d_2)^2$ and is inversely proportional to $(\cos ({\theta _t})\cos ({\theta _r}))$. Then in logarithmic scale, it can be readily inferred that the PL factor on $d_1$ (or $d_2$) is -2 and the PL factor on $\cos ({\theta _t})$ (or $\cos ({\theta _r})$) is -1 respectively. However, considering the non-linearity and irreversibility between discrete phase shifts and continuous phase shifts, the practical PL factors in discrete space are eager to be clarified, which may be of great significance in many RIS-related fields such as channel modeling. In this subsection, we provide the PL scaling law under discrete phase shifts designed by DTPQ, which could serve as the upper bound due to the optimality of DTPQ.

\begin{figure}[htbp]
\vspace{-0.5cm}
\centering
\begin{minipage}[t]{0.48\textwidth}
\centering
\includegraphics[width=3in]{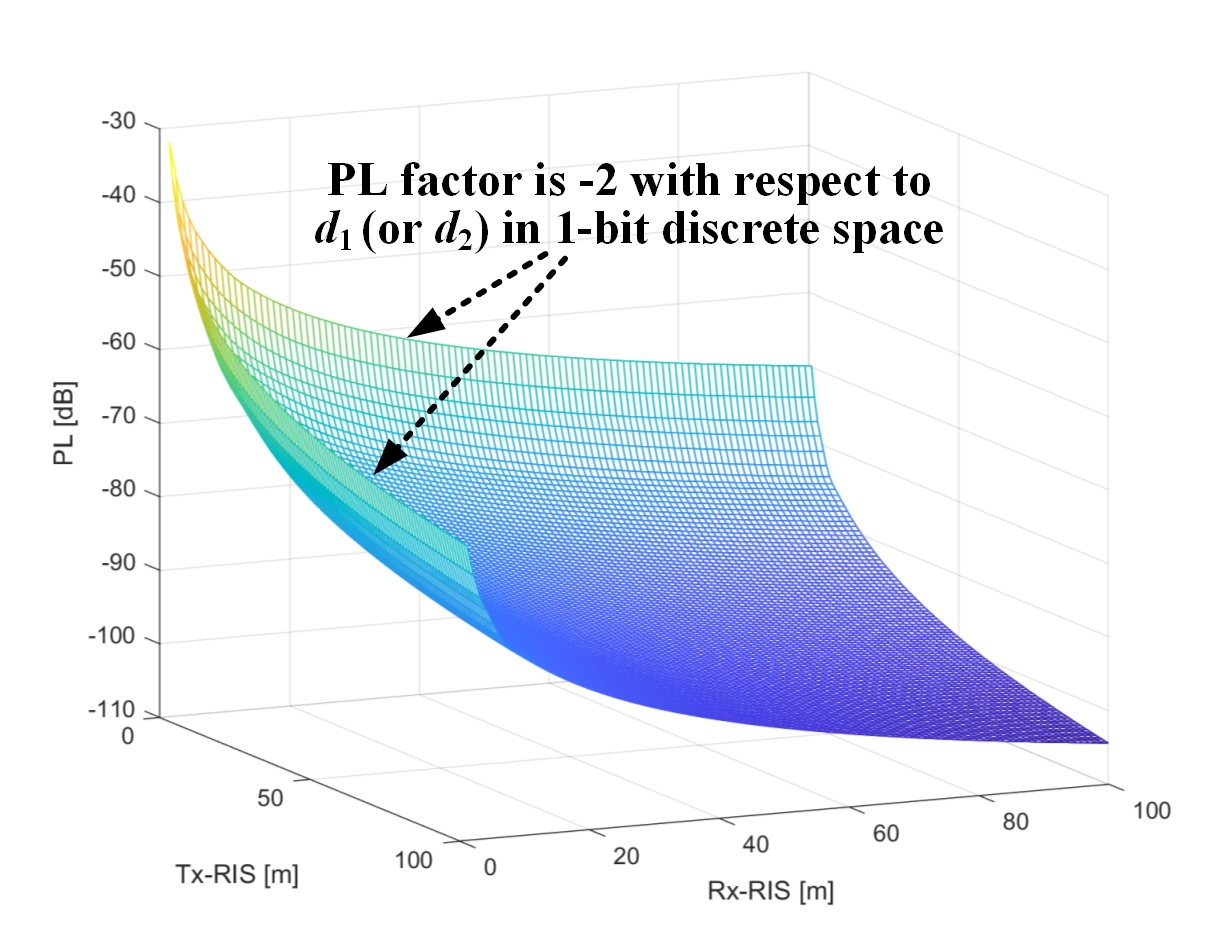}
\vspace{-0.5cm}
\caption{PL scaling law v.s. $d_1$ and $d_2$.}
\label{fig_PLlaw}
\end{minipage}
\begin{minipage}[t]{0.48\textwidth}
\centering
\includegraphics[width=3in]{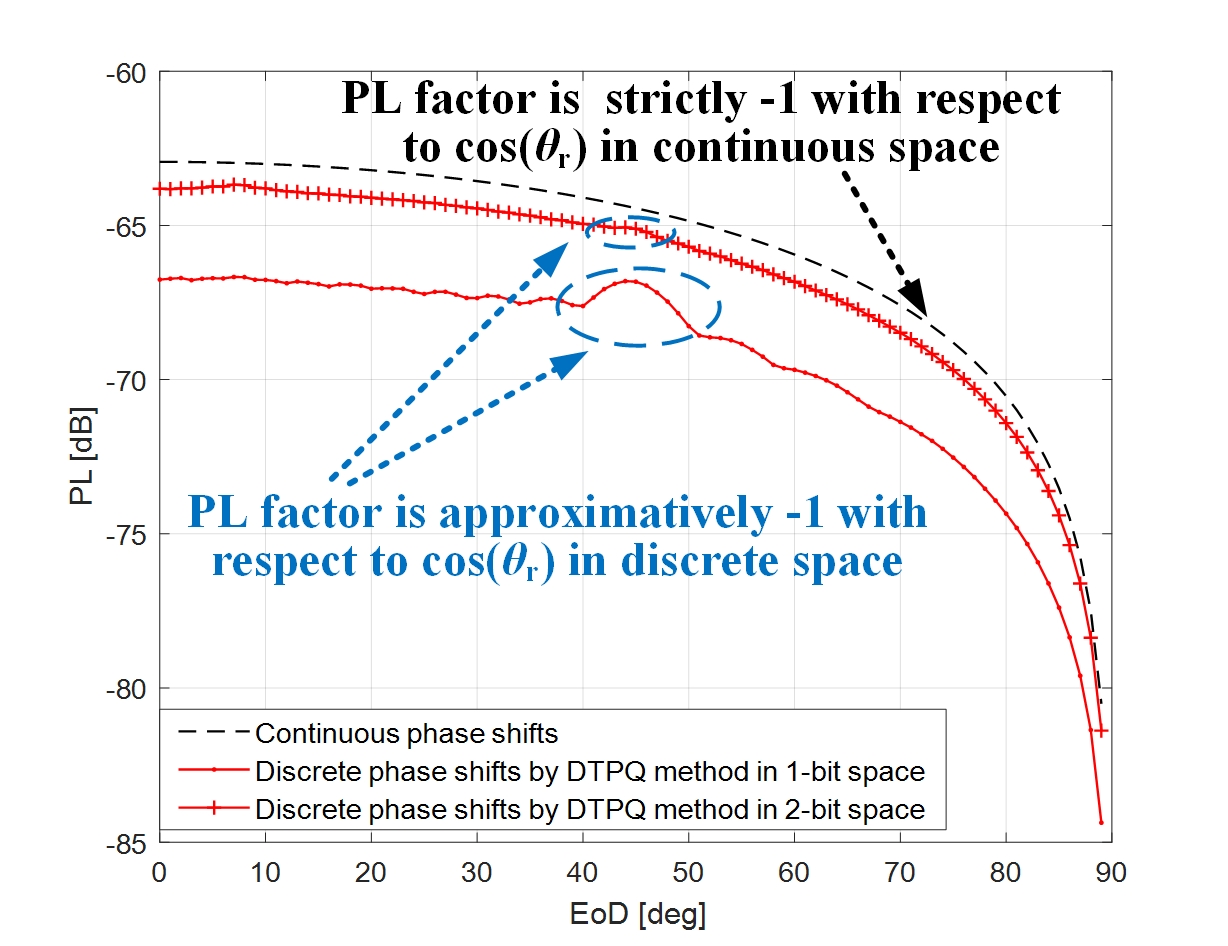}
\vspace{-0.5cm}
\caption{PL scaling law v.s. $\theta_r$.}
\label{fig_anglelaw}
\end{minipage}
\vspace{-0.5cm}
\end{figure}

\reffig{fig_PLlaw} and \reffig{fig_anglelaw} reveal the PL scaling laws on distance and angle in discrete space respectively, where \textbf{RIS 1} in \autoref{table2} is utilized for analyses. In detail, \reffig{fig_PLlaw} exhibits the PL scaling laws with respect to $d_1$ and $d_2$ in 1-bit discrete space. In this figure, the PL factor on $d_1$ (or $d_2$) is -2 in logarithmic scale. This phenomenon indicates that the PL scaling law is still proportional to $(d_1d_2)^2$ in discrete space under the phase shifts designed by DTPQ. It agrees well with the PL characteristics in continuous space described in \cite{b5}.

\reffig{fig_anglelaw} illustrates the PL scaling laws on EoD $\theta_r$ in continuous space, in 1-bit discrete space and in 2-bit discrete space respectively. The phase shift levels of \{55° 235°\} in 1-bit discrete space and the phase shift levels of \{55° 145° 235° 325°\} in 2-bit discrete space are considered. Both of the discrete phase shift matrix of 1-bit and that of 2-bit are designed by DTPQ. As shown in this figure, the PL factor on $\cos(\theta_r)$ in discrete space is not strictly -1 in logarithmic domain, which shows a ``crest'' (``blue circle'' in this figure) where the EoD $\theta_r$ equals to the EoA $\theta_t$ ($\theta_r$=$\theta_t$=45° in this figure). This phenomenon is slightly different from that observed in continuous space (``black dotted line'' marked in this figure). This phenomenon can be explained by the property of discrete phase shifts that signal energy is better focused, when Tx and Rx locate at the specular positions with respect to RIS. In addition, we observed that the ``crest'' is most conspicuous in 1-bit discrete space. As the number of bits $q$ increases, the ``crest'' smooths gradually and the PL is also improved. As $q$ tends to infinity, discrete phase alignment tend to be perfect continuous phase alignment. It is worth mentioning that the similar PL scaling law on EoA $\theta_t$ can be inferred due to the reciprocity of $\theta_t$ and $\theta_r$.

\subsection{Field trials}

\begin{figure*}[htbp]
     \centering
       \subfloat[]{\includegraphics[height = 2.4in]{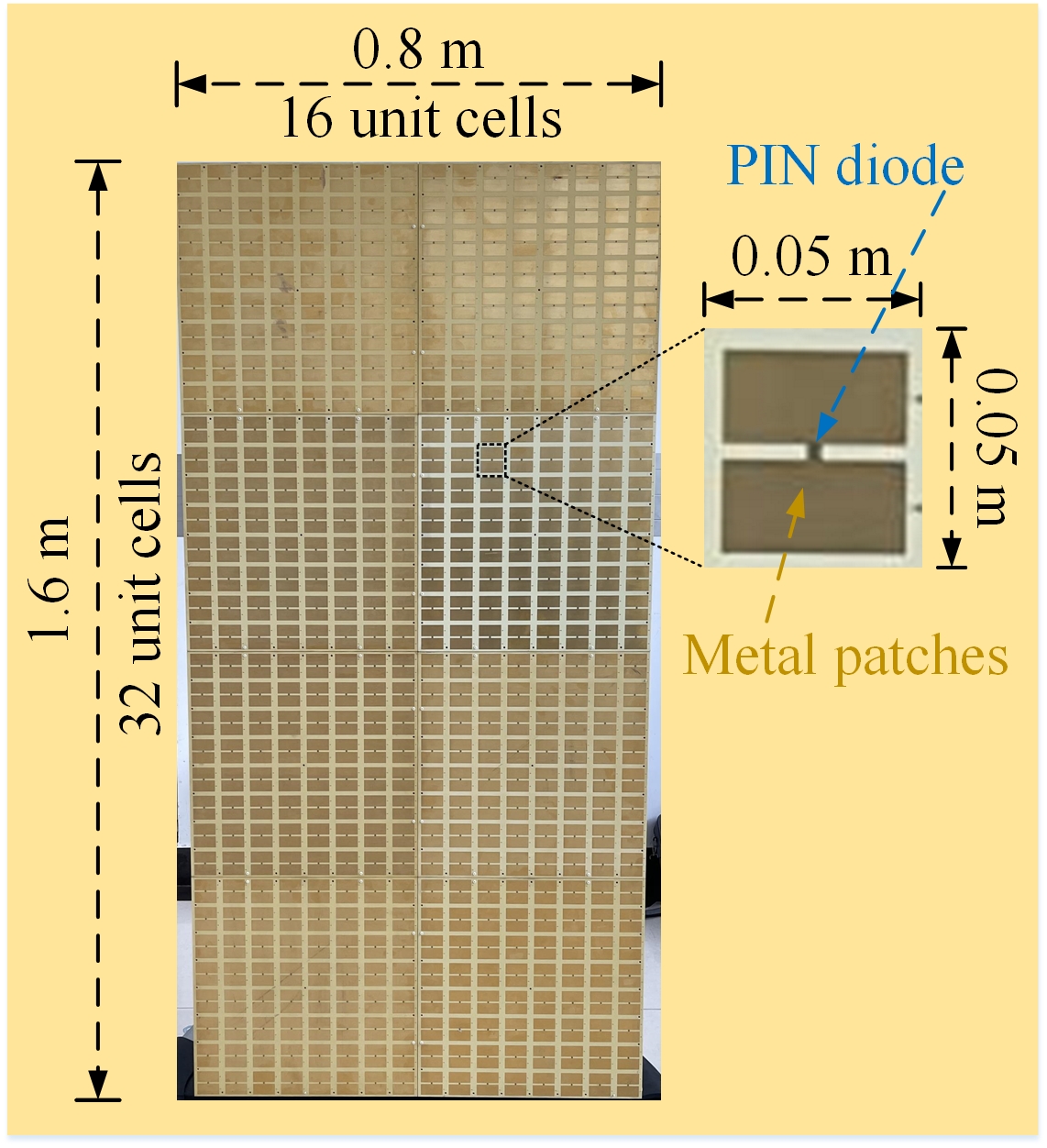}\label{2.6GRIS}} \hspace{1cm} 
       \subfloat[]{\includegraphics[height = 2.4in]{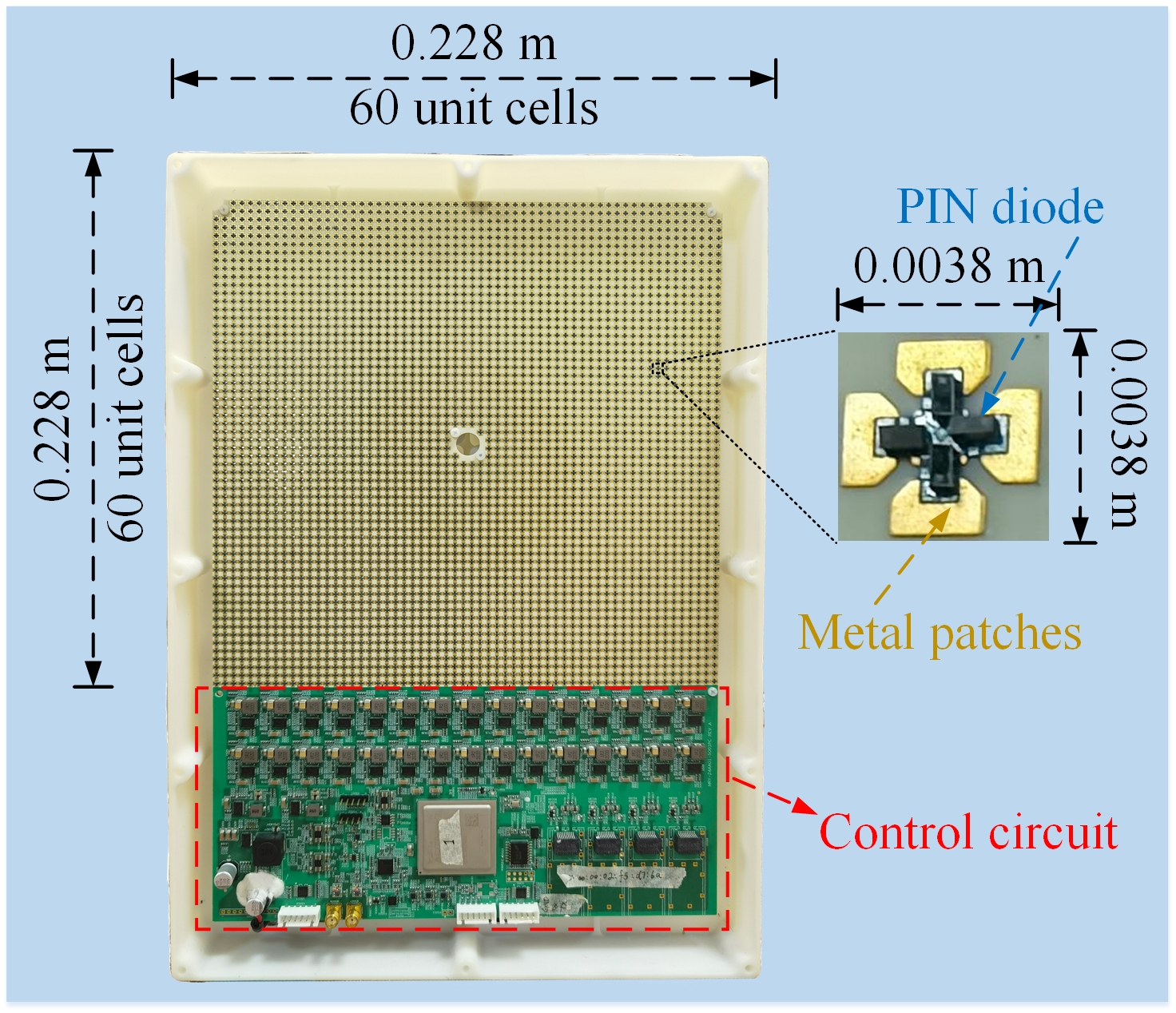}\label{35GRIS}} 
    \caption{Images of RISs. (a) 2.6 GHz RIS. (b) 35 GHz RIS. }
\label{RISpicture}
\vspace{-0.5cm}
\end{figure*}

To validate the application performance of the proposed method in practical environment, two field trials are carried out at sub-6 GHz and mmWave frequency bands respectively, in Jiulonghu Campus of Southeast University, Nanjing, China. An outdoor open area is selected as trial site. Two RISs working at 2.6 GHz and 35 GHz respectively, are deployed in the corresponding field trials, which are shown in \reffig{RISpicture}. \reffig{2.6GRIS} illustrates the RIS at the center frequency of 2.6 GHz, which is composed of 32 unit cells per column and 16 unit cells per row. The unit cell on this RIS has a size of 0.05 m $\times$ 0.05 m. \reffig{35GRIS} exhibits the RIS at the center frequency of 35 GHz, which is constituted of 60 unit cells per column and 60 unit cells per row. The unit cell on this RIS has a size of 0.0038 m $\times$ 0.0038 m. The unit cells of these two RISs are comprised of metal patches and positive-intrinsic-negative (PIN) diodes, which are of 1-bit phase resolution and independently programmed. 

\begin{figure*}[htbp]
     \centering
       \subfloat[]{\includegraphics[height = 2.5in]{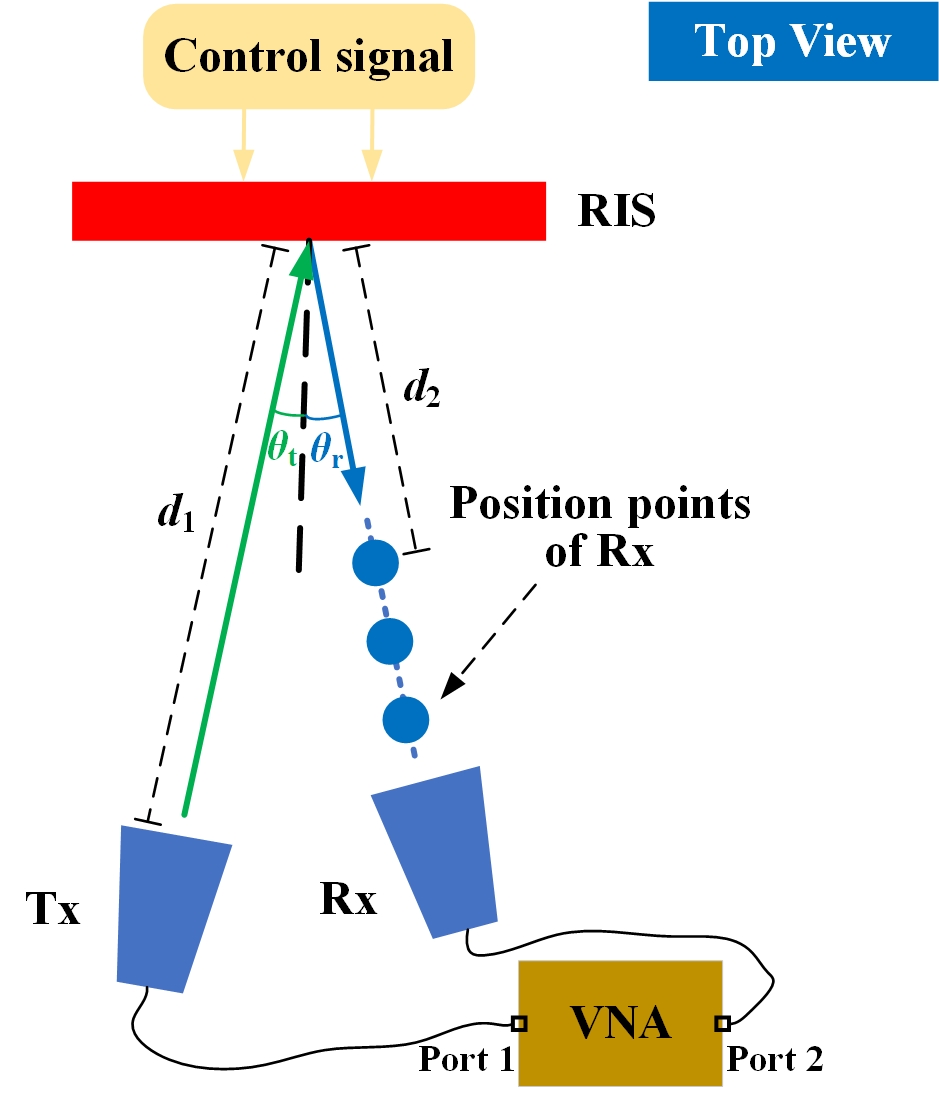}\label{plan}} \hspace{2cm} 
       \subfloat[]{\includegraphics[height = 2.5in]{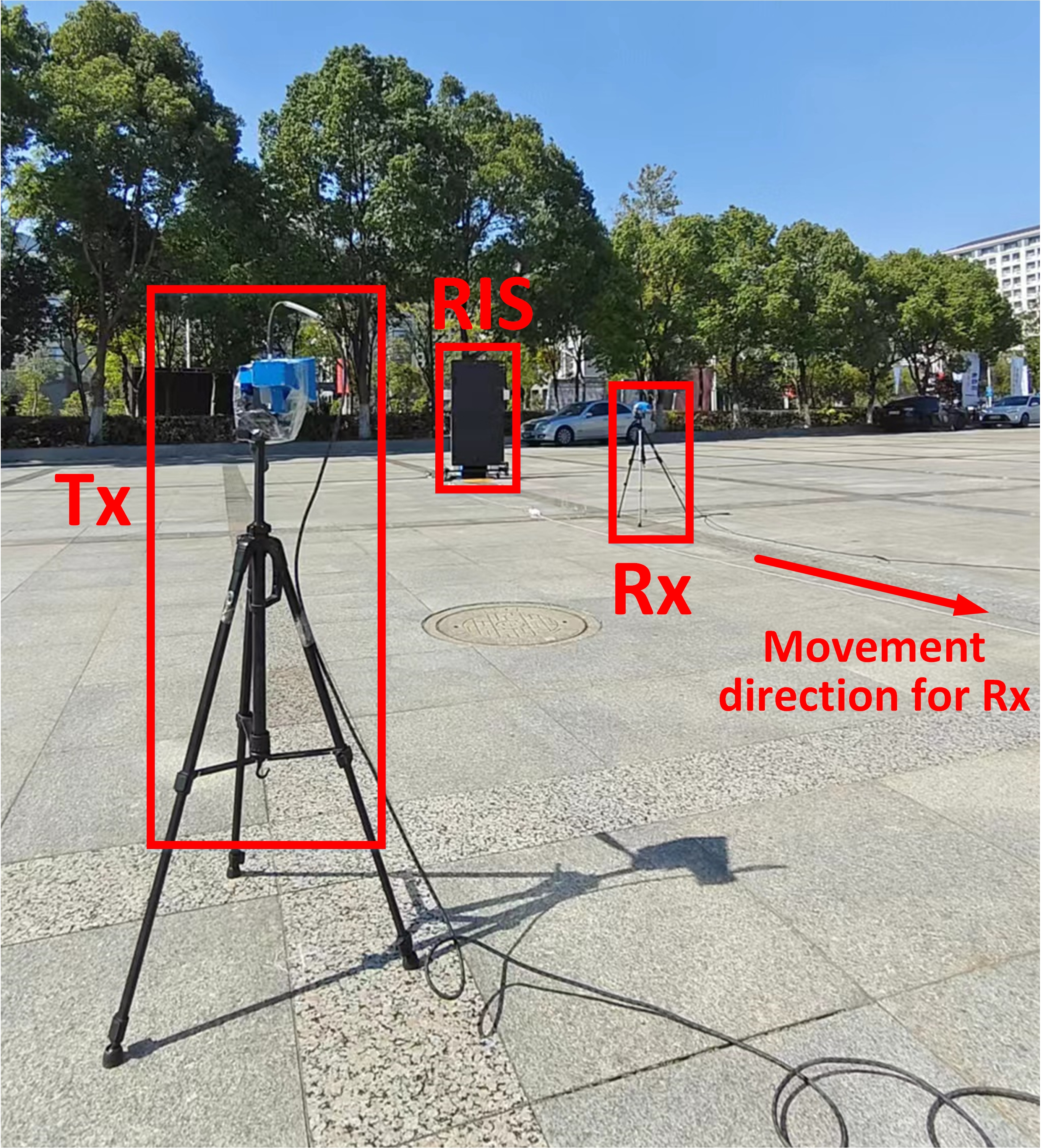}\label{fig_fieldpic}} 
    \caption{Descriptions of field trials. (a) Schematic plans. (b) Images. }
\label{field trials}
\vspace{-0.5cm}
\end{figure*}

The schematic plan of field trials is visually illustrated in \reffig{plan}. The Ceyear vector network analyzer (VNA) with operating frequency ranged from 10 MHz to 67 GHz is utilized to collect the trial data. Two horn antennas acting as Tx and Rx respectively, are connected to port 1 and port 2 of VNA, which are placed facing the deployed RIS with EoA $\theta_t$ = EoD $\theta_r$ = 10°. These horn antennas are maintained at a same height as the center of RIS, thus the AoA $\varphi_t$ = 0° and the AoD $\varphi_r$ = 180° can be assumed. Due to the fact that the horn antennas of Tx and Rx face nearly the same direction, the direct propagation link between Tx and Rx is considered to be blocked and the dominated propagation is the RIS-assisted cascaded channel. The image of field trials is illustrated in \reffig{fig_fieldpic}. The detailed information on the deployed RISs and measurement system configurations in these two field trials are described in \autoref{table3}.

\begin{table}[htbp]
\vspace{-0.5cm}
\caption{Parameters for field trials}
\label{table3}
    \centering
\begin{tabular}{|c|c|c|}
\hline
\textbf{Parameter}              & \textbf{RIS 1}     & \textbf{RIS 2}       \\ \hline
Operating frequency $f$         & 2.6 GHz            & 35 GHz               \\ \hline
Wavelength $\lambda$            & 11.5 cm            & 8.6 mm             \\ \hline
Number of rows $M$              & 32                 & 60                   \\ \hline
Number of columns $N$           & 16                 & 60                   \\ \hline
Size of unit cell               & 5 cm$\times$5 cm   & 3.8 mm$\times$3.8 mm  \\ \hline
Phase resolution   & 1-bit              & 1-bit                \\ \hline
Discrete phase shift levels     & $\rho_1$=55°, $\rho_2$=235°  & $\rho_1$=0°, $\rho_2$=180°     \\ \hline
Tx-RIS distance $d_1$           & 10 m               & 10 m                 \\ \hline
Rx-RIS distance $d_2$           & 5:0.1:10 m         & 5:0.1:10 m           \\ \hline
EoA $\theta_t$ & 10°             & 10°                  \\ \hline
EoD $\theta_r$ & 10°             & 10°                  \\ \hline
AoA $\varphi_t$  & 0°          & 0°                     \\ \hline
AoD $\varphi_r$  & 180°        & 180°                   \\ \hline
Antenna gain of Tx    &  8.25 dBi   &  25.62 dBi        \\ \hline
Antenna gain of Rx    &  8.25 dBi   &   25.62 dBi       \\ \hline
HPBW of Tx antenna    & 60°    & 8°                  \\ \hline
HPBW of Rx antenna    & 60°    & 8°                  \\ \hline
Transmitted power of VNA  & 0 dBm     & 0 dBm     \\ \hline
Average factor of VNA           & 5      & 5            \\ \hline
Measured bandwidth         & 10 MHz    & 10 MHz   \\ \hline
Number of scanning frequency points & 101 & 101 \\ \hline
\end{tabular}
\vspace{-0.5cm}
\end{table}

\begin{figure}[htbp]
\vspace{-0.5cm}
\centering
\begin{minipage}[t]{0.48\textwidth}
\centering
\includegraphics[width=3in]{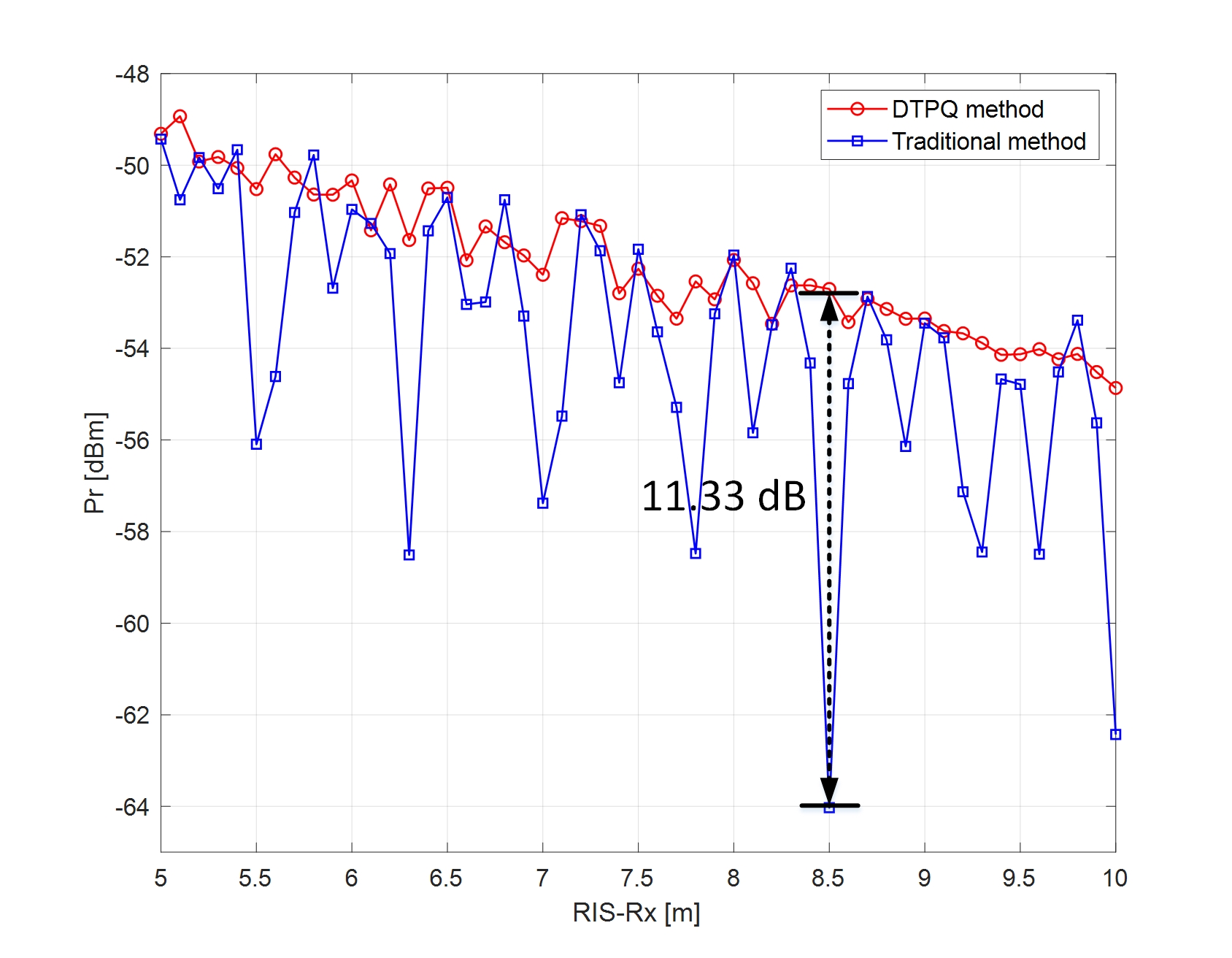}
\vspace{-0.5cm}
\caption{Trial results with 2.6 GHz RIS.}
\label{trial_2.6G}
\end{minipage}
\begin{minipage}[t]{0.48\textwidth}
\centering
\includegraphics[width=3in]{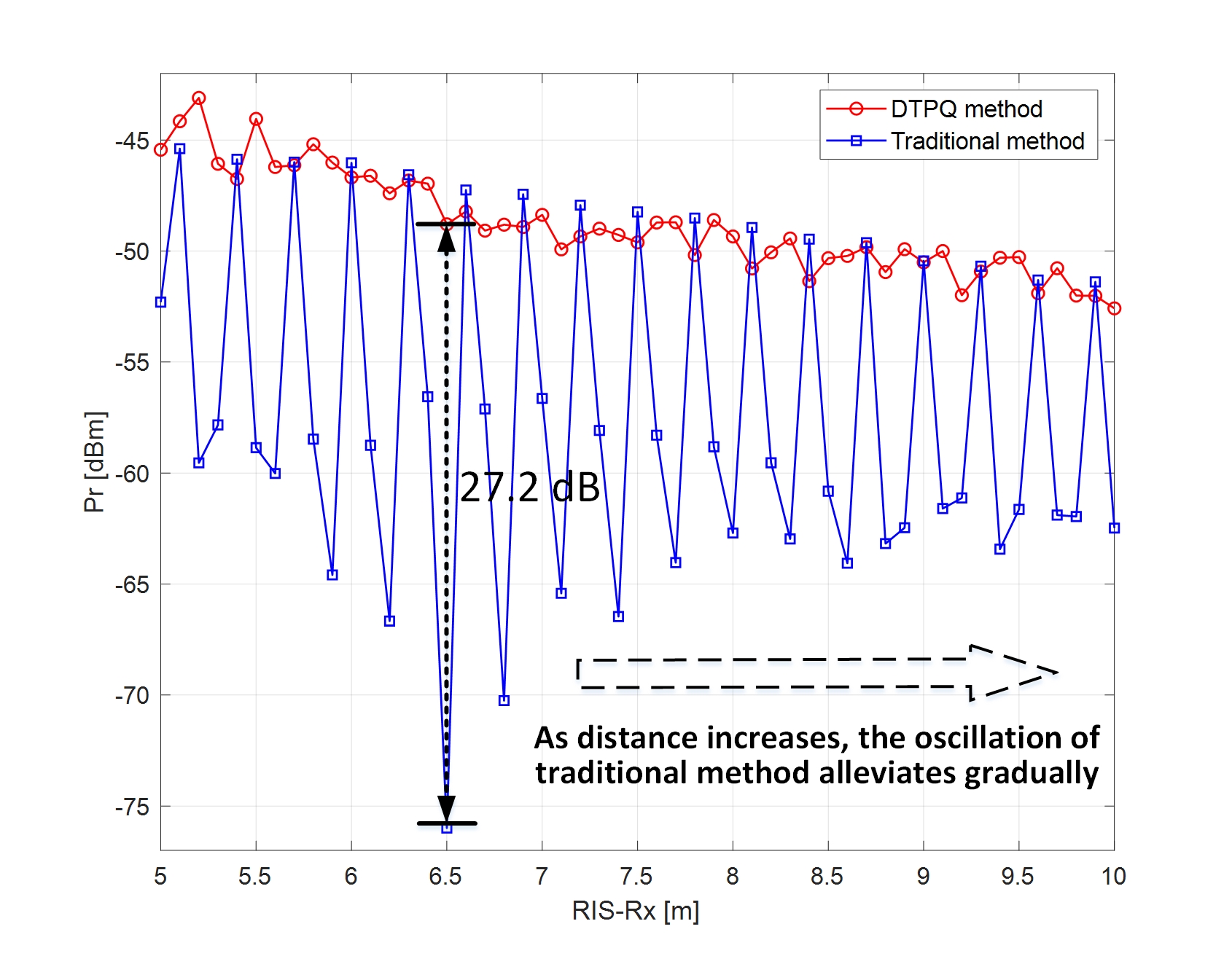}
\vspace{-0.5cm}
\caption{Trial results with 35 GHz RIS.}
\label{trial_35G}
\end{minipage}
\end{figure}

\reffig{trial_2.6G} illustrates the trial results at 2.6 GHz with \textbf{RIS 1} in \autoref{table3} applied. The ``blue line'' in this figure represents the received power under discrete phase shifts designed by traditional method, which adopts the 235° ($\rho_2$ of \textbf{RIS 1} in \autoref{table3}) as the fixed quantization threshold. It shows severe oscillations performance, which is in accordance with the theoretical analyses previously. Meanwhile, the ``red line'' denotes the received power under discrete phase shifts designed by DTPQ method, which manifests a stable and significant improvement, i.e., the maximum power gain of about 11.33 dB, against traditional method. 

\reffig{trial_35G} shows the trial results at 35 GHz with \textbf{RIS 2} in \autoref{table3} deployed. The traditional method in this figure utilizes 180° ($\rho_2$ of \textbf{RIS 2} in \autoref{table3}) as the fixed quantization threshold. In \reffig{trial_35G}, the DTPQ method still performs better than traditional method and demonstrates a maximum power gain achieving even 27.2 dB against traditional method. In addition, two notable phenomena can be observed. Firstly, the ``oscillation'' of traditional method becomes more intense compared to \reffig{trial_2.6G}. This phenomenon may result from that the wavelength at 35 GHz is much shorter than that at 2.6 GHz, thus the phase periodicity at 35 GHz is more dense within the trial distance of 5$\sim$10 m. Secondly, as the distance increases, the ``oscillation'' of traditional method is alleviated gradually. It agrees well with our analyses in Section III that ``traditional quantization method may cause a quantization error, \textit{especially in the near field}''. As a result, both of \reffig{trial_2.6G} and \reffig{trial_35G} validate the significant improvements on power gain as well as stability of the proposed DTPQ method against those of the traditional method in practical communication environment.

\section{Conclusions}
\label{sec5}
In this paper, we investigated the phase alignment and quantization for discrete phase shift designs in RIS-assisted SISO system. Firstly, we analyzed the characteristics of phase distribution in the near field and far field of RIS respectively, which could seriously affect the discretization of phase shift for RIS. Aiming to maximize beamforming effect by aligning phases, we unveiled the phase distribution law and its underlying DoF, serving as the guideline of phase quantization strategies. We showed that the optimal discrete phase shifts of RIS can be quantized from its counterparts in continuous space, with the appropriate quantization threshold set. Applying such analytical result, we proposed two phase quantization methods, i.e., DTPQ and EIPQ. The proposed DTPQ method could calculate the optimal discrete phase shift matrix to align phases of EM waves at Rx, with complexity linearly dependent on the number of unit cells on RIS. Whilst, the proposed EIPQ method could obtain a sub-optimal solution yet with an even lower complexity of constant, compared to DTPQ method. Simulation results indicated that both of the proposed methods achieved significant improvements on received power, stability, as well as robustness, against the traditional method. The PL scaling laws in discrete space was also unveiled under the discrete phase shifts designed by DTPQ due to its optimality, which might promote the RIS-related researches such as channel modeling. Two field trials at sub-6 GHz and mmWave frequency bands were carried out respectively in practical communication environment to validate the potential applications of the proposed methods.

{\appendix[Proof of Theorem 1 by reduction to absurdity]

\begin{figure*}[htpb]
\vspace{-0.5cm}
     \centering
       \subfloat[]{\includegraphics[scale = 0.26]{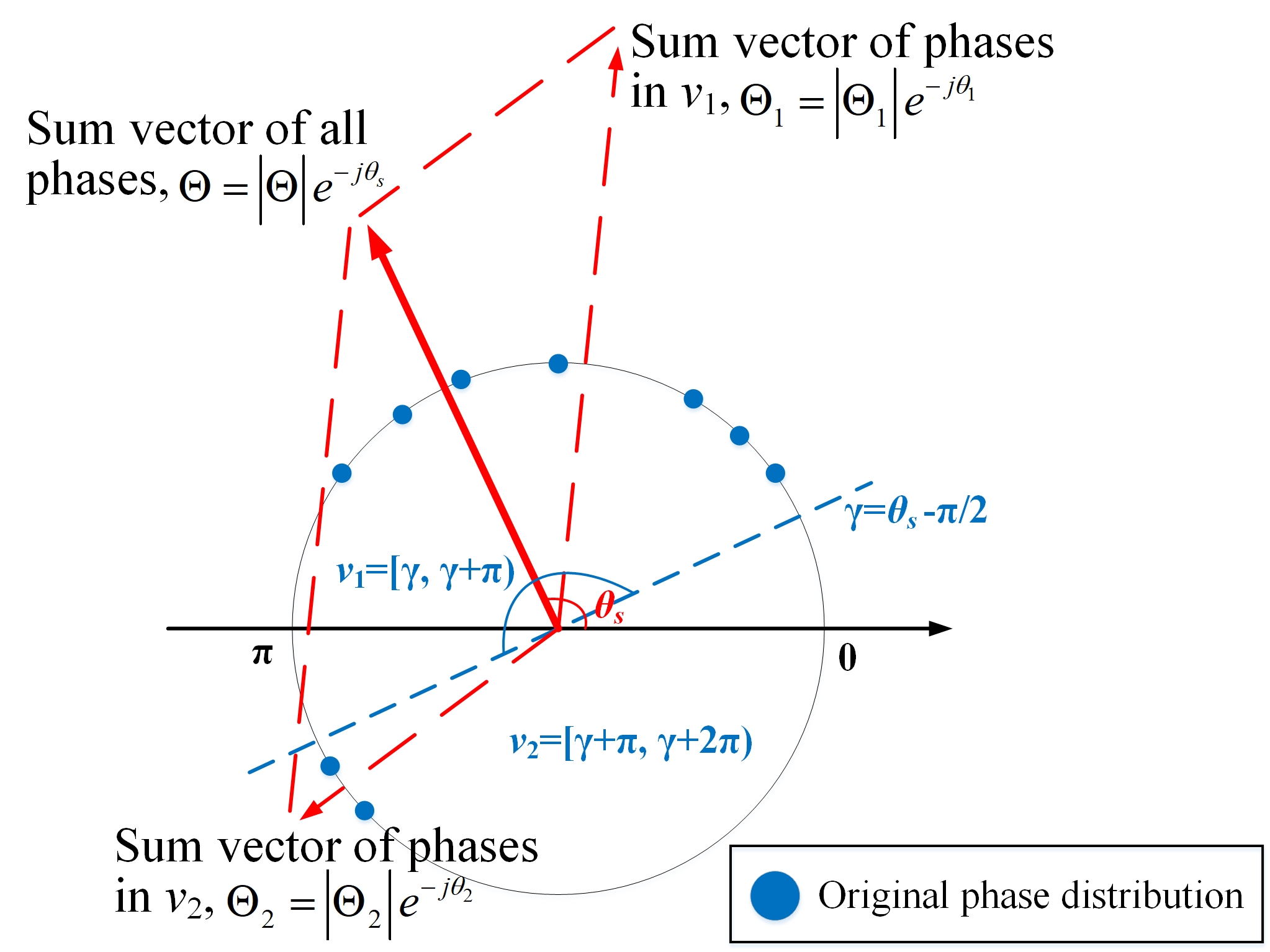}\label{proofa}} \hspace{1cm}
       \subfloat[]{\includegraphics[scale = 0.26]{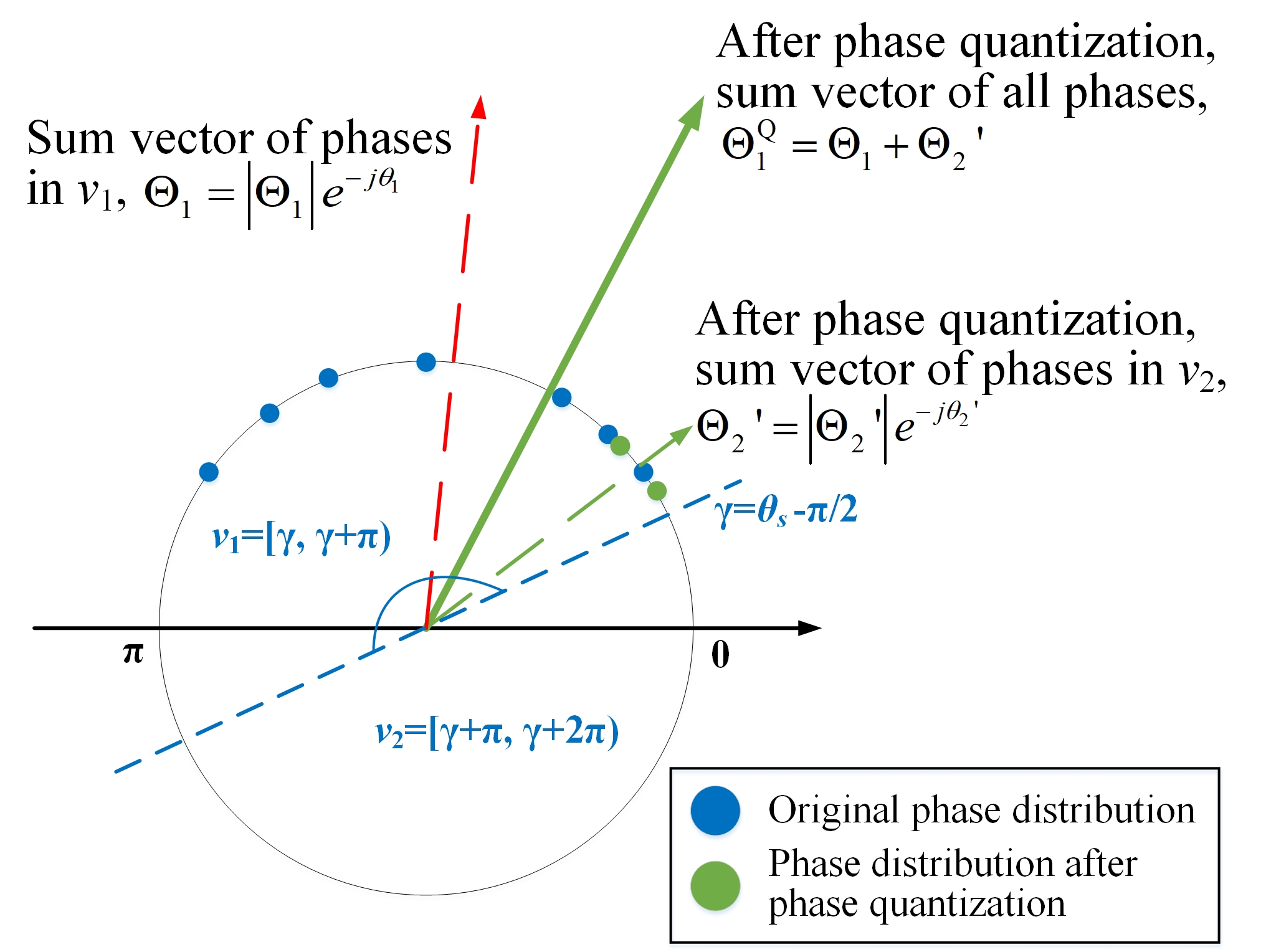}\label{proofb}} 
    \caption{Illustrations for proof in 1-bit discrete space. (a) Original phase distribution of EM waves. (b) Phase distribution of EM waves after phase quantization. }
\label{proof}
\vspace{-0.5cm}
\end{figure*}

According to \eqref{eq2}, given the EM waves at Rx denoted as ${\bm{{\rm X}}}{e^{( - j{\bf{\Phi }})}}$, where ${\bm{{\rm X}}}$ is the magnitude matrix of these EM waves. The element located at the $m$th column and the $n$th row of ${\bm{{\rm X}}}$ can be expressed as ${{\rm X}_{n,m}} = \sqrt {F_{n,m}^{combine}} /(r_{n,m}^tr_{n,m}^r)$ . Note that ${\bm{{\rm X}}}$ is constant matrix  when the positions of Tx, RIS and Rx are fixed. ${\bf{\Phi }}$ denotes the phase matrix of these EM waves, which is equivalent to continuous phase shift matrix of RIS and can be calculated according to \eqref{eq_r5}. The sum vector of ${\bm{{\rm X}}}{e^{( - j{\bf{\Phi }})}}$ is denoted as $\vec \Theta  = \left| {\vec \Theta } \right|{e^{ - j{\theta _s}}}$, with its phase denoted as ${\theta _s}$ in polar coordinates. Through configuring the appropriate phase shifts on RIS, the phases of EM waves at Rx could be aligned to enhance their sum vector $\vec \Theta $. 

We make the proof of Theorem 1 by contradiction.

\textit{Assumption 1: Assume there exists the optimal phase distribution of EM waves that maximizes phase alignment but violates \eqref{eq_r6}.}

\textcircled {1} Firstly, we analyze the above assumption for the case of 1-bit discrete space with phase shift levels of 0 and $\pi$. As shown in \reffig{proofa}, consider phase quantization threshold ${\gamma} = {\theta_s} - \pi/2$. Then the quantization range can be divided into ${\nu _1} = [{\gamma} , \ {\gamma} + \pi )$ and ${\nu _2} = [{\gamma} + \pi, \ {\gamma} + 2\pi)$. The sum vector of the phases inside ${\nu _1}$ is denoted as ${\vec \Theta _1}$, and the sum vector of the phases inside ${\nu _2}$ is denoted as ${\vec \Theta _2}$. It is clear that $\vec \Theta  = {\vec \Theta _1} + {\vec \Theta _2}$. Correspondingly, ${\theta _1} \in [{\gamma}, \ {\gamma} + \pi)$ and ${\theta _2} \in [{\gamma} + \pi, \ {\gamma} + 2\pi)$ represent the phases of $\vec \Theta _1$ and $\vec \Theta _2$, respectively.

Through configuring phase shifts of $\pi$ (i.e., phase quantization), the phases in ${\nu _2}$ could be turned into ${\nu _1}$ and $\vec \Theta _2$ will be correspondingly transformed into ${\vec \Theta _2}'$, which are illustrated as \reffig{proofb}. It can be easily deduced that the phase of ${\vec \Theta _2}'$ can be written as ${\theta _{{2}}}' = {\theta _{{2}}} + \pi  \in [\gamma,\ \gamma + \pi)$ and its magnitude can be expressed as $\left| {{{\vec \Theta }_2}'} \right| = \left| {{{\vec \Theta }_2}} \right|$. Assume that the sum vector of phases for all the EM waves after phase quantization as $\vec \Theta _1^{\rm{Q}}$ and $\vec \Theta _1^{\rm{Q}} = {\vec \Theta _1} + {\vec \Theta _2}'$ can be inferred. Then, the magnitudes of $\vec \Theta$ and $\vec \Theta _1^{\rm{Q}}$ can be expressed as \eqref{11a} and \eqref{11b} respectively.
\begin{figure*}[htbp]
\vspace{-0.5cm}
\begin{subequations}
\begin{align}
\left| {\vec \Theta } \right|{\rm{ }} = \left| {\sqrt {{{({{\vec \Theta }_1} + {{\vec \Theta }_2})}^2}} } \right| = \sqrt {\left| {{{({{\vec \Theta }_1})}^2}} \right| + \left| {{{({{\vec \Theta }_2})}^2}} \right| + 2\left| {{{\vec \Theta }_1}} \right|\left| {{{\vec \Theta }_2}} \right|\cos ({\theta _1} - {\theta _2})},  \label{11a}\\
\left| {\vec \Theta _1^{\rm{Q}}} \right| = \left| {\sqrt {{{({{\vec \Theta }_1} + {{\vec \Theta }_2}')}^2}} } \right| = \sqrt {\left| {{{({{\vec \Theta }_1})}^2}} \right| + \left| {{{({{\vec \Theta }_2}')}^2}} \right| + 2\left| {{{\vec \Theta }_1}} \right|\left| {{{\vec \Theta }_2}'} \right|\cos ({\theta _1} - {\theta _2}')}.  \label{11b}
\end{align}
\end{subequations}
\vspace{-0.5cm}
\end{figure*}

Then, by comparing the magnitudes of $\vec \Theta$ and $\vec \Theta _1^{\rm{Q}}$, we can verify that whether \textit{Assumption 1} is correct or false. We divide ${\theta _1} \in [{\gamma}, \ {\gamma} + \pi)$ into two cases including {\em case 1}: ${\theta _1} \in [\gamma+\pi/2,\ \gamma+\pi)$ and \em case 2}: ${\theta _1} \in [\gamma,\ \gamma +\pi/2)$. For {\em case 1} that ${\theta _1} \in [\gamma+\pi/2,\ \gamma+\pi)$, ${\theta _2} \in [\gamma + 3\pi/2,\ \gamma + 2\pi)$ and ${\theta _2}' \in [\gamma+\pi/2,\ \gamma+\pi)$ can be inferred according to the dot product theorem of vectors. Consequently, $\pi > \left| {{\theta _1} - {\theta _2}} \right| > \left| {{\theta _1} - {\theta _2}'} \right| > 0$ and $\cos (\left| {{\theta _1} - {\theta _2}} \right|) < \cos (\left| {{\theta _1} - {\theta _2}'} \right|)$ can be deduced. Note that for {\em case 2} that ${\theta _1} \in [\gamma,\ \gamma +\pi/2)$, we can obtain ${\theta _2} \in [\gamma +\pi,\ \gamma +3\pi/2 )$ and ${\theta _2}' \in [\gamma,\ \gamma+\pi/2)$. Then $\cos (\left| {{\theta _1} - {\theta _2}} \right|) < \cos (\left| {{\theta _1} - {\theta _2}'} \right|)$ still holds. As a consequence, $\left| {\vec \Theta } \right| < \left| {\vec \Theta _1^{\rm{Q}}} \right|$, indicating the phase distribution of EM waves in \textit{Assumption 1} is not optimal. Thus, in 1-bit discrete space, the above \textit{Assumption 1} is invalid and \textbf{Theorem 1} holds.
 
\textcircled {2} Secondly, in 2-bit discrete space, the above-mentioned ${\nu _1}$ and ${\nu _2}$ can be extended to ${\nu _1}=[{\gamma}, \ {\gamma} + 2\pi /{2^2})$, ${\nu _2}=[{\gamma} + 2\pi /{2^2}, \ {\gamma} + 4\pi /{2^2})$, ${\nu _3}=[{\gamma} + 4\pi /{2^2}, \ {\gamma} + 6\pi /{2^2})$, ${\nu _4}=[{\gamma} + 6\pi /{2^2}, \ {\gamma} + 8\pi /{2^2})$. Through configuring appropriate phase shifts within $\{0 \ \pi/2 \ \pi \ 3\pi/2 \}$ (i.e., phase  quantization), the phases in ${\nu _2}$, ${\nu _3}$ and ${\nu _4}$ could be turned into ${\nu _1}$. Denote the sum vector of phases for all the EM waves after phase quantization as ${\vec \Theta _2^{\rm{Q}}}$. It is easily deduced that $\left| {\vec \Theta _2^{\rm{Q}}} \right| > \left| {\vec \Theta _1^{\rm{Q}}} \right| > \left| {\vec \Theta } \right|$, which can be proved similarly to the procedure of \textcircled {1}. Thus, \textit{Assumption 1} is still invalid and \textbf{Theorem 1} still holds in 2-bit discrete space. Similarly, we can further infer that $\left| {\vec \Theta _q^{\rm{Q}}} \right| >  \ldots  > \left| {\vec \Theta _2^{\rm{Q}}} \right| > \left| {\vec \Theta _1^{\rm{Q}}} \right| > \left| {\vec \Theta } \right|$, which indicates that \textit{Assumption 1} is always invalid in discrete space with arbitrary number of bits. Thus, in order to maximize phase alignment in $q$-bit discrete space, \eqref{eq_r6} should be satisfied, that is, any two quantized phases should be within a quantization interval of $\Omega=2\pi /{2^q}$.

\textbf{Theorem 1} is thus proved.



\vfill

\end{document}